\documentclass[epj]{svjour}

\usepackage{graphics}
\usepackage{epsfig}

\journalname{The European Physical Journal B}

\newcommand{\be}{\begin{equation}}
\newcommand{\ee}{\end{equation}}
\newcommand{\bea}{\begin{eqnarray}}
\newcommand{\eea}{\end{eqnarray}}
\newcommand{\bfig}{\begin{figure}}
\newcommand{\efig}{\end{figure}}
\newcommand{\bc}{\begin{center}}
\newcommand{\ec}{\end{center}}
\newcommand{\btab}{\begin{tabular}}
\newcommand{\etab}{\end{tabular}}
\newcommand{\dr}{\partial}

\let\oldepsilon\epsilon
\let\epsilon\varepsilon
\let\varepsilon\oldepsilon
\let\oldphi\phi
\let\phi\varphi
\let\varphi\oldphi
\def\EQ{\begin{equation}}
\def\EN{\end{equation}}
\def\EQA{\begin{eqnarray}}
\def\ENA{\end{eqnarray}}

\newcommand{\ddx}{\partial_{x}}

\newcommand{\ccc}{$\mbox{\textsf{C}}_{\mbox{\textsf{c}}}^{\mbox{\textsf{c}}}$}

\begin{document}

\title{Selection of dune shapes and velocities.\\
Part 2: A two-dimensional modelling.}

\author{
Bruno Andreotti\inst{1}
\and
Philippe Claudin\inst{2}
}

\institute{
Laboratoire de Physique Statistique de l'Ecole Normale Sup\'erieure,
24 rue Lhomond, 75231 Paris Cedex 05, France.
\and
Laboratoire des Milieux D\'esordonn\'es et H\'et\'erog\`enes (UMR 7603),
4 place Jussieu - case 86, 75252 Paris Cedex 05, France.
}

\date{\today}

\abstract{ We present in this paper a simplification of the dune model
proposed by Sauermann \textit{et al}.  which keeps the basic
mechanisms but allows analytical and parametric studies.  Two kinds
of purely propagative two dimensional solutions are exhibited: dunes
and domes, which, by contrast to the former, do not show avalanche
slip face.  Their shape and velocity can be predicted as a function of
their size.  We recover in particular that dune profiles are not scale
invariant (small dunes are flatter than the large ones), and that the
inverse of the velocity grows almost linearly with the dune size.  We
furthermore get the existence of a critical mass below which no stable
dune exists.  However, the linear stability analysis of a flat sand
sheet shows that it is unstable at large wavelengths and suggests a
mechanism of dune initiation.   }

\authorrunning{B. Andreotti and P. Claudin}
\titlerunning{Selection of dune shapes and velocities. Part 2}
\maketitle

\section{Introduction}
The beauty of the crescentic barchan dunes have recently attracted the
interest of physicists for a better understanding and modelling of
sand transport, as well as ripples and dunes formation and
propagation.  E. Guyon had this witty remark about them: `barchans are
our drosophila', which means that beyond the scientific and
fundamental works on these dunes, we all keep in mind that such
studies may lead to potential applications in the fight of saharan
countries against sand invasion.  One of the first reference work in
the field is certainly the famous book of Bagnold which dates back
from 1941 \cite{B41}.  Since then, a great effort of measurement and
modelling has been done which we have reviewed in details in the first
part of these twin papers.

Our aim here is to discuss and model the selection of two-dimensional
dune shape and velocity.  For that purpose, we will simplify the model
proposed by Sauermann \textit{et al}.  \cite{S01,SKH01,KSH01}.  We
will show that although rather severe approximations, we are able to
recover their main results, in particular that dune profiles are not
scale invariant, and that the inverse of the velocity grows almost
linearly with the dune size.  Besides, analytical expressions of dome
and dune propagative profiles can be obtained, but whose coefficients
have to be numerically computed.  We furthermore get the existence of
a critical mass below which no stable dune exists.  The apparition of
dunes can however be understood with the linear stability analysis of
a flat sand sheet which is unconditionally unstable towards large
wavelengths perturbations.

The paper is organized as follows.  Section \ref{equasdebase} is
devoted to the equations of the model.  The linear stability of a
uniform sand sheet is treated in section \ref{stability}.  In section
\ref{cccmodel} we simplify further the equations and show what is the
general shape of the purely propagative solutions of the model.  The
specific case of domes and `actual' dunes are discussed in sections
\ref{domes} and \ref{dunes} respectively.  At last, we conclude with a
discussion of the relevance of these results, and the possible
extension of the model to three-dimensional situations and dynamical
studies.

\section{Basic equations}
\label{equasdebase}
We wish to give a description of the shape and evolution of two-dimensional
dunes in terms of two fields: the profile $h(x,t)$ and the volumic sand
flux $q(x,t)$ which is the volume of sand transported through an
infinite vertical line per unit time.  $x$ denotes the horizontal
coordinate and $t$ is the time.  We are going to write a set of three
equations for these quantities in order to include into the model (i) the
mass conservation, (ii) the progressive saturation of sand transport and
(iii) the feedback of the topography on the sand erosion/deposition processes.
Although we shall keep in this paper to two-dimensional situations
which correspond to transverse dunes (invariant in the direction
perpendicular to the wind), our ultimate goal is of course to be able
to describe three dimensional dunes and barchans in particular.

Because barchans in dune fields -- so-called `ergs' -- organize themselves
like gooses or ducks during their migration flights, we named this
class of models for purely graphical reasons the \ccc\ modellings. This
denomination includes the approach of Sauermann \textit{et al}. as
well as the different variations and simplifications we derived in
this paper from their work. At this stage of the modelling, we are however
far from being able to take into account such `interactions' between dunes
which are necessary to explain the \ccc\ spatial organisation, and we shall
focus on isolated objects only.

A simple balance calculation shows that the erosion rate $-\dr_t h$ is
directly related to the divergence of the flux $q$.  This gives the
common continuity equation:
\be
\label{equacont}
\dr_t h + \dr_x q = 0.
\ee

\bfig[t]
\bc
\epsfxsize=\linewidth
\epsfbox{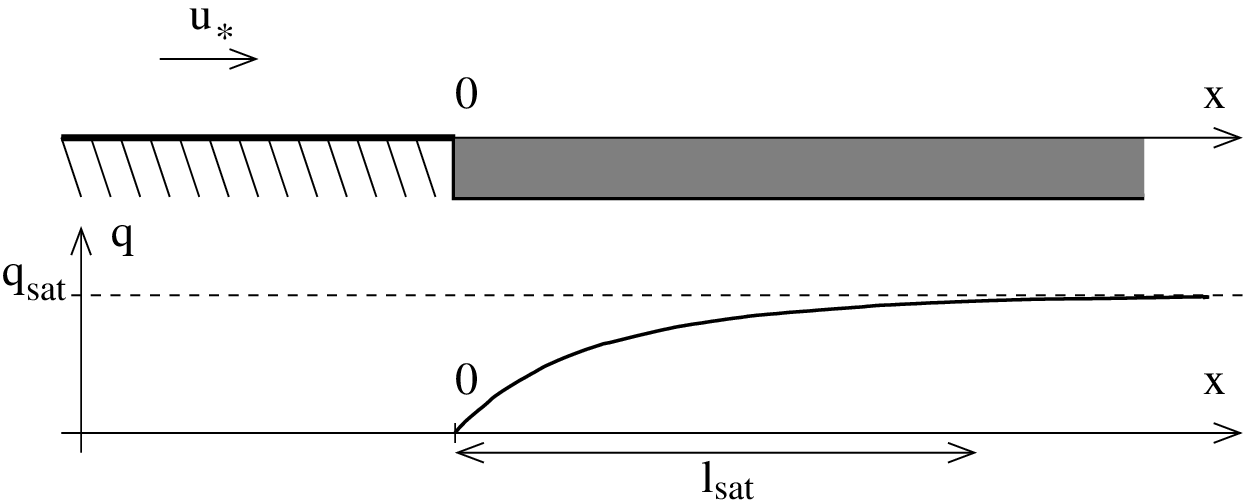}
\caption{When the wind is blowing on a patch of sand, the flux of transported
sand get saturated after a typical length $l_{sat}$ which is almost
independent of the wind shear velocity $u_*$ -- see part 1.}
\label{charge}
\ec
\efig

The saturation effect of the sand transport has been already evoked in
the first part of these twin papers: consider a patch of sand on which
the wind is blowing -- see figure \ref{charge}.  The flux of
transported sand $q$ first increases and, because of the feedback of
the grains on the wind velocity profile, get saturated after a typical
length $l_{sat}$.  In the first part of the paper, we showed that
$l_{sat} = \xi d \, \rho_{sand}/\rho_{air}$ ($d$ is the grain
diameter, $\rho$ the densities and $\xi$ a non dimensional prefactor),
i.e.  is almost independent of the wind shear velocity $u_*$ -- slow
logarithmic dependencies hidden in $\xi$ only.  This phenomenon has
been reported and studied by several authors, e.g.  \cite{B41,AH88}.
The real shape of $q(x)$ is certainly more complicated than the one
drawn on figure \ref{charge}.  In particular, oscillating or
overshooting features were reported in \cite{B41} when $q$ reaches its
asymptote.  However, what is important for our purpose is only that a
saturated value $q_{sat}$ is reached after a length $l_{sat}$.  This
space lag is satisfactorily described by the following equation:
\be
\label{equacharge}
\dr_x q = \frac{q_{sat} -q}{l_{sat}}.
\ee

This charge equation can be seen as a simplification of that proposed
by Sauermann \textit{et al}. in their continuum saltation model \cite{SKH01}.
An important remark is that this equation is valid only if \emph{some
grains are available on the sand bed}.  On a firm soil indeed, the
flux cannot increase to become saturated.  As suggested by Peer and
Hakim \cite{PH01}, the right hand side of the relation
(\ref{equacharge}) must be therefore multiplied by some matching function
which quickly tends to zero when $h$ is decreased below, say, $d$ the grain
diameter, and which is equal to unity above this value -- the altitude
of the firm soil is $h=0$.  Then, the equation (\ref{equacharge})
becomes non linear but no boundary conditions have to be specified at
the edges of the sand covered region.  For example, if the matching function
tends to zero like $h$, the dune will always keep a thin sand sheet at its
back. But if it varies as $\sqrt{h}$, the dune will have a finite
extension and will join the firm soil with an horizontal tangent.  We
shall ignore at present these subtleties but keep them in mind to
invoke them later when necessary.

Another important remark is that the time scale on which the dune
profile $h$ evolves is incomparably larger than that of the sand flux
$q$.  We then assume that $q$ adapts its profile instantaneously
according to equation (\ref{equacharge}) and makes $h$ change slowly
through equation (\ref{equacont}).  Therefore any term $\dr_t q$ is
irrelevant in this modelling.

The saturated flux $q_{sat}$ is uniform for a flat sand bed only.  To
the first order, the saturated flux $q_{sat}$ is a function of the
local shear stress $\tau=\rho_{air} u_*^2$ which itself depends -- non
locally -- on the topography: basically, bumps and upwind slopes get
more eroded than dips and downwind faces.  A classical relationship
between the saturated flux and the shear velocity that can be
recovered with the scaling arguments of the part 1 of the paper is:
\be
q_{sat} \propto \frac{\rho_{air}}{\rho_{sand}} \frac{u_*^3}{g}.
\ee
In principle, such a relationship is valid far from the velocity
threshold $u_{thr}$ under which no sand can be eroded by the wind,
i.e. $q_{sat}=0$ for $u_* \le u_{thr}$. Refined formulas can be obtained
which essentially smooth the step from $0$ to the previous asymptotic
expression such as that obtained in part 1:
\be
q_{sat} \propto \frac{\rho_{air}}{\rho_{sand}} \frac{u_*}{g}
\left(u_*^2-u_{thr}^2\right).
\ee
In the whole range $-u_{thr}<u_*<u_{thr}$, $q_{sat}$ is null so that
in practice, $q_{sat}$ cannot become negative on a dune.  This
condition will be used in section \ref{domes}.

To close the equations, we have to explicit the spatial variations of the
turbulent wind velocity due to the dune profile.  The simplest model
which verifies the basic requirements (see part 1) is certainly the
perturbative calculation by Jackson and Hunt \cite{JH75,Weng91}.
Neglecting logarithmic scale dependencies, Kroy \textit{et al}. \cite{KSH01}
have extracted the main features of their work by expressing the shear
velocity as:
\be
\label{convolu2}
\frac{u_*^2(x)}{U_*^2} = 1 + A \int \!\!  \frac{ds}{\pi s} \, \dr_x
h(x-s) + B \, \dr_x h(x),
\ee
where $U_*$ is the shear velocity exerted on a flat bed. More precisely,
as discussed in details in the part 1, $A$ and $B$ in principle depend on
the size of the dune $D$ with $\ln D/z_0$ factors, where $z_0$ is the
roughness of the sand surface. For $D$ varying between $20$ and $200~m$
and a roughness of order of the grain size, such a logaritmic factor does
not change by more than $20\%$ over the whole range. This justifies the fact
that it is reasonable to take constant effective values for the
coefficients $A$ and $B$. Several further important remarks must be made
on this expression.  First, it must be noted that the convolution integral
acts on $\ddx h$ roughly like a derivative, leading to a term which encodes
curvature effects.  But this curvature is dimensionless and thus does not
depend on the dune size -- in other words, this term can be seen as a
curvature rescaled by the dune size.  It reflects the observation that
the wind velocity increases on bumps (negative curvature) and decreases
on hollows (positive curvature).  Second, it is a non local term, meaning
that the shear velocity depends on the whole shape of the dune. Of course
sharp variations of the dune profile will also have a strong local
effect.  At last, the second term simply takes into account slope
effects: positive slopes are more eroded than negative ones.  Again,
this term does not introduce any new lengthscale.

Expression (\ref{convolu2}) can be used to close up the set of equations
as was done by Kroy \textit{et al}.  in \cite{KSH01} who used besides
a more sophisticated -- and non-linear -- charge equation than
(\ref{equacharge}).  It is useful to simplify further the equation
linking $q_{sat}$ to $h$ without loosing too much physics, in order to
let more analytical developments.  The linear expansion of the
expression (\ref{convolu2}) rewritten in terms of $q_{sat} \propto
u_*^3$ gives:
\be
\label{convolqsat}
\frac{q_{sat}(x)}{Q_{sat}} = 1 + \frac{3}{2} A \int \!\!
\frac{ds}{\pi s} \, \dr_x h(x-s) + \frac{3}{2} B \, \dr_x h(x),
\ee
where $Q_{sat}=q_{sat}(U_*)$ is the saturated flux on a flat sand bed
submitted to a shear velocity $U_*$.

Furthermore, we shall use the saturation length $l_{sat}$ and flux
$Q_{sat}$ to make our variables dimensionless.  Thus, for a given wind
shear velocity, all relevant scales of the problem are fixed.  For
instance, $Q_{sat}/l_{sat}$ is the velocity scale, $l_{sat}^2/Q_{sat}$
the time scale, $Q_{sat}/l_{sat}^2$ the frequency scale, etc. Note that
the strength of the wind is completely encoded in these rescalings.

\section{Stability of a flat sand bed}
\label{stability}
Before going further in the modelling of dune shape and propagation,
we wish to investigate the problem of dune initiation.  Indeed, there
are two striking field observations.  First, no persistent barchan
dunes exist smaller than say, $1~m$ high, $20~m$ large and $20~m$
long.  Second, any small conical sandpile blown by the wind disappears
even when a sand supply is provided.  Then, how can barchan dunes
appear?

\bfig[t]
\bc
\epsfxsize=\linewidth
\epsfbox{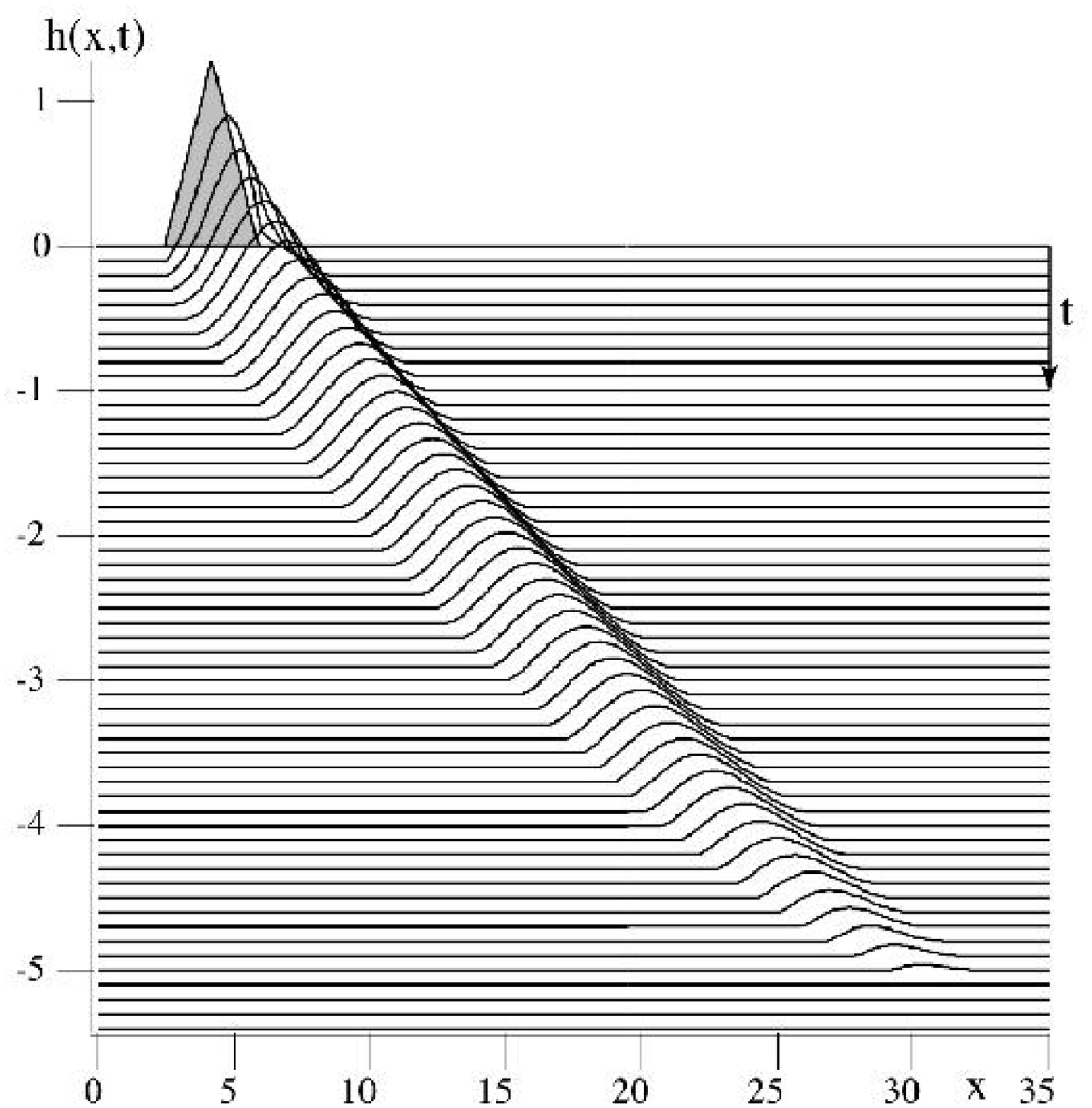}
\caption{Numerical integration of equations
(\ref{equacont},\ref{equacharge},\ref{convolqsat}) for the evolution of a
small triangular sandpile on the firm soil, initially $1.3~l_{sat}$ high and
at the repose angle. $h(x,t)$ and $x$ are in units of $l_{sat}$. The time
between two profiles is $0.1$ in units of $l_{sat}^2/Q_{sat}$. For
legibility, the profiles are translated vertically from time to time.
The grey filled region shows the available amount of sand at $t=0$. The
field observation that such a small sandpile disappears when blown by the
wind is then recovered in the model.}
\label{Triangle}
\ec
\efig
\bfig[t]
\bc
\epsfxsize=\linewidth
\epsfbox{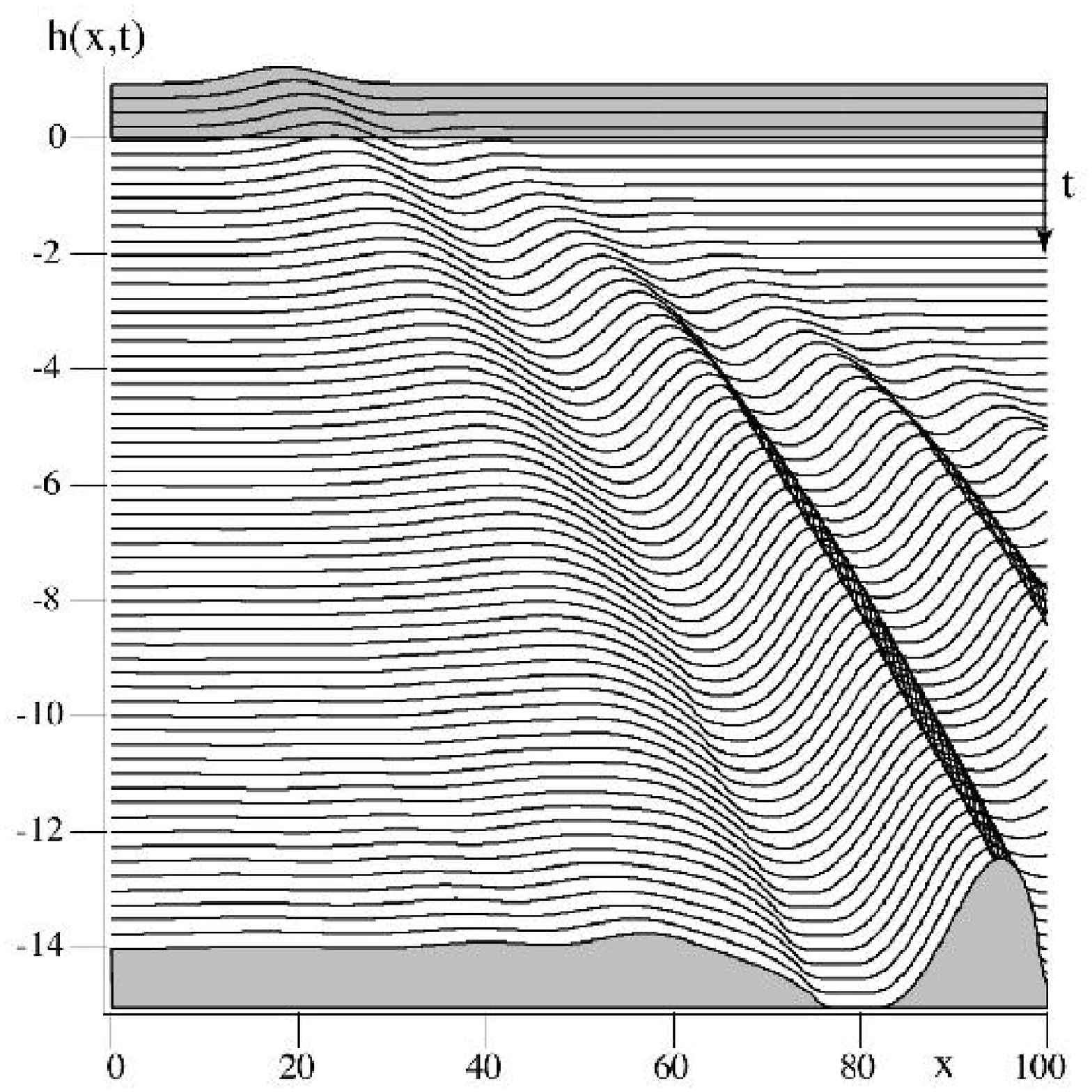}
\caption{Numerical integration of equations
(\ref{equacont},\ref{equacharge},\ref{convolqsat}) for the evolution of a flat
sand sheet of width $l_{sat}$ over the firm soil disturbed by a small
hump. $h(x,t)$ and $x$ are in units of $l_{sat}$. The time between two
profiles is $0.5$ in units of $l_{sat}^2/Q_{sat}$. For legibility, the
profiles are translated vertically from time to time. The grey filled
region shows the available amount of sand at $t=0$ and at the final time.
The perturbation propagates downwind and growing oscillations nucleate from
it. The hollow created in front of the initial bump soon reaches the
firm soil.}
\label{Plage}
\ec
\efig

To investigate this problem, we have integrated numerically equations
(\ref{equacont},\ref{equacharge},\ref{convolqsat}) with $A=6$ and
$B=4$.  Two initial conditions were tested.  First, a small triangular
sandpile at the repose angle is prepared on the firm soil
(figure~\ref{Triangle}).  It can be seen that it is rapidly eroded and
disappears, as observed in the field and in wind tunnel experiments --
see part 1.  Second, we look at the evolution of a thin sand sheet
(figure~\ref{Plage}) disturbed by a flat bump.  This initial
conditions mimics a sand beach on which sand is deposed by water.  It
can be seen that the bump propagates downwind and induces a strong
erosion of the sand bed in front of it.  A second bump nucleates from
the initial perturbation which itself induces a strong erosion in
front of it, and so on.  After some time, a series of growing
oscillations is generated.  The amplification of this phenomenon stops
when the oscillations eventually reach the firm soil from which no
sand can be eroded, and/or when recirculation bubbles -- see section
\ref{dunes} -- appear and make these bumps interact.  As a conclusion,
depending on their spatial extension, small sand bumps either
disappear or grow and initiate dunes.

It is then instructive to make the linear stability analysis of a
flat sand bed.  Let us consider an infinite uniform sand bed blown by
a uniform wind.  The sand flux is everywhere saturated: $q=1$ (in
units of $Q_{sat}$).  To investigate its stability, we can consider,
without loss of generality a small perturbation of the form:
\bea
h(x,t) & = & H e^{\sigma t - i \omega t + i k x}, \\
q(x,t) & = & 1 + Q e^{\sigma t - i \omega t + i k x},
\eea
where $Q$ and $H$ are related one to the other by the conservation of matter,
\be
\label{stab1}
(i \omega-\sigma) H= i k Q.
\ee
>From the relation (\ref{convolqsat}), we get the following expression
for the saturated flux: $q_{sat}=1 + 3/2 (|k|A + ikB) h$.  Once
replaced in the saturation equation, it gives:
\be
\label{stab2}
(1+i k l_{sat}) Q = \frac{3}{2} (A |k|l_{sat}+B i k) H.
\ee
Combining equations (\ref{stab1}) and (\ref{stab2}) we finally obtain:
\bea
\label{sigma}
\sigma & = & \frac{3 k^2 (B-A |k|)}{2 (1+k^2)}, \\
\label{omega}
\omega & = & \frac{3 k|k|(A+B |k|)}{2 (1+k^2)}.
\eea
Note that a more complicated relation between $q_{sat}$ and $u_*$
would only affect the prefactor, i.e.  the time scale, in this
calculation.

\bfig[t]
\bc
\epsfxsize=\linewidth
\epsfbox{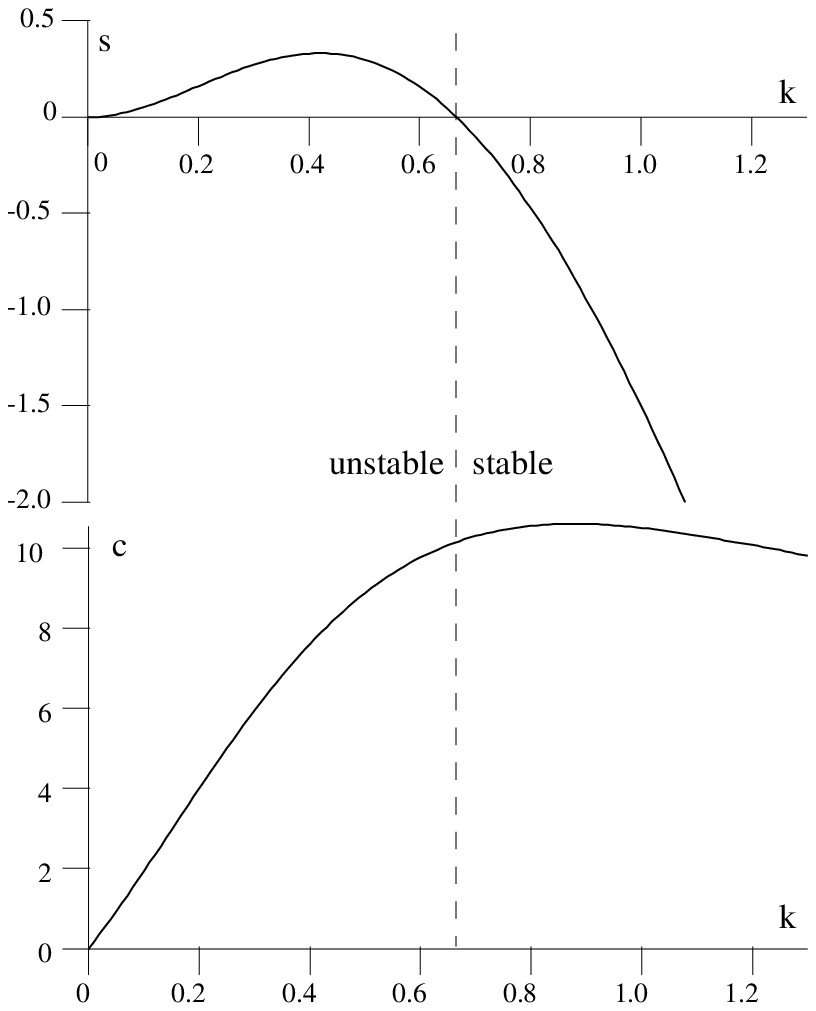}
\caption{Results of the linear stability of a uniform sand bed blown
by the wind.  Top: growth rate $\sigma$, rescaled by
$Q_{sat}/l_{sat}^2$, as a function of the disturbance wave number $k$,
rescaled by $1/l_{sat}$.  The sand bed exhibits a large wavelength
instability.  Bottom: group velocity $c$ of disturbances, rescaled by
$Q_{sat}/l_{sat}$, as a function of the disturbance wave number $k$,
rescaled by $1/l_{sat}$. For small wave numbers, the velocity $c$
increases linearly with $k$ and thus decreases as the inverse of the
wavelength.}
\label{LinStaAn}
\ec
\efig

The growth rate $\sigma$, shown for $A=6$ and $B=4$ on
figure~\ref{LinStaAn} (top), is positive for small wavenumbers
($k<B/A$) and negative for large wavenumbers ($k>B/A$).  A flat sand
bed thus exhibits a large wavelength instability which can explain the
initiation of dunes.  But it is stable towards small wavelength
disturbances, so that a small sandpile on a firm soil is quickly
eroded.

The fastest growing mode -- that which maximises $\sigma$ -- is
obtained for $3k+k^3=2B/A$.  Neglecting the $k^3$ term, we obtain a
still good approximation of the most unstable wavelength:
$\lambda=2\pi/k \simeq 3\pi l_{sat} A/B$.  For $A=6$ and $B=4$ we get
$\lambda \simeq 14 \, l_{sat}$.  For $l_{sat} \simeq 9~m$ we get a
wavelength which is reasonable compared to what is observed on
transverse dune fields in deserts.  Under water $l_{sat}$ rather
scales on the grain size and we have $l_{sat} \simeq 1~cm$ which gives
again a good estimation of what has been measured by Betat \textit{et
al}.  \cite{BKFR01} or Andersen \textit{et al}.  \cite{AGRL01} for
example.

Finally, it can be seen from equation (\ref{omega}) that disturbances
propagate downwind.  We can compute the group velocity of these
surface waves:
\be
\label{velostaana}
c = \frac{d\omega}{dk}=\frac{3 [A |k| + B k^2 (3 +k^2)]}{2 (1+k^2)^2}.
\ee
$c$ is plotted on figure~\ref{LinStaAn} (bottom) for $A=6$ and $B=4$. For
small wavenumbers $k$, the group velocity $c$ increases linearly with
$k$.  This means that for asymptotically large wavelengths $\lambda$,
the propagation speed scales as $Q_{sat}/\lambda$. We thus recover in
the limit of large bumps, the scaling proposed by Bagnold. We also
see that the velocity deviates significantly from this law, when the
wavelength becomes comparable to the saturation length $l_{sat}$. This
is also the case for actual dunes -- see part 1.

\section{A simple 2d modelling}
\label{cccmodel}

\subsection{Simplified equations}

The expression (\ref{convolqsat}) can be kept as it is, mixed with the
mass conservation (\ref{equacont}) and the charge (\ref{equacharge})
equations, and treated in Fourier space, but the idea is rather to
replace the convolution expression by a simpler anzats.  Since it is a
`scaleless and non-local curvature', we can replace it by $-D \dr_{xx} h$,
where $D$ is the dune length.  $D$ will be determined by the boundary
conditions.  We then end up with the following set of -- linear --
equations of our model:
\bea
\label{ccc1}
\dr_t h + \dr_x q & = & 0, \\
\label{ccc2}
\dr_x q & = & q_{sat} -q, \\
\label{ccc3}
q_{sat} & = & 1 - \alpha D \dr_{xx} h + \beta \dr_x h.
\eea
$\alpha$ and $\beta$ are two of the parameters of the model -- a third
one, $\mu_b$, will be introduced later on. They could be in principle
measured independently. The $\alpha$ term has a stabilizing role on small
perturbations, while the $\beta$ one make them growing. The rescalings
$h \leftarrow \alpha h$ and $t \leftarrow \alpha t$ (with $q$ and $x$
unchanged) shows that only the ratio $\beta/\alpha$ is actually important.
As stated in the figure captions, most of the curves of the paper have
been plotted with $\alpha=1.$ and $\beta=4$. At last, as we already
mentioned, the charge equation (\ref{ccc2}) is valid only if there is some
sand to be eroded $h>0$.  Note besides that, to be fully consistent,
$q_{sat}$ should remain positive everywhere.

\subsection{Boundary conditions}

Before getting into the process of solving this system of equations,
we need to specify what the boundary conditions at the edges of the
dune are. This rises several important questions. The first
thing to notice is that, when we transformed the relation
(\ref{convolqsat}) which links $q_{sat}$ to the profile $h$ with a
convolution term into the simpler expression (\ref{ccc3}), we have
incremented the order of the differential equation.  Therefore, such a
simplification requires an additional boundary condition than what is
needed to integrate equations (\ref{convolqsat},\ref{ccc1},\ref{ccc2})
for example. First order equations like the later need to know
the profile $h$ at some position. However, when using the second order
equations (\ref{ccc1}-\ref{ccc3}), the slope $h'$ should be also specified.
As we already mentioned, one way to solve this problem is to regularise
equation (\ref{ccc2}) by a non linear factor when $h$ tends to zero which
gives conditions on $h(x)$ and $h'(x)$ as $x \to -\infty$, see discussion
after equation (\ref{equacharge}) in section \ref{equasdebase}.
\bfig[t!]  \bc \epsfxsize=\linewidth \epsfbox{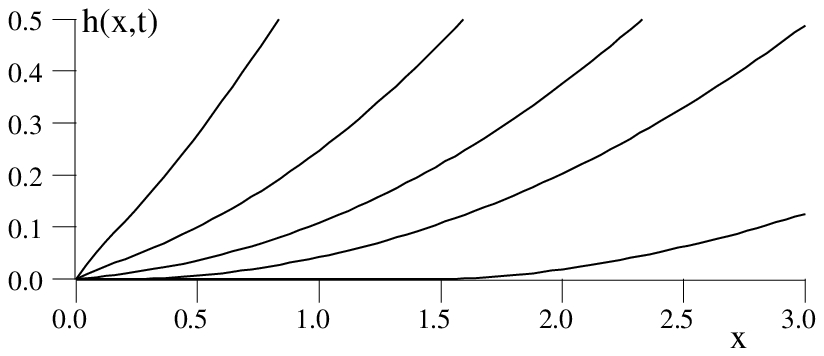}
\caption{The numerical integration of equations
(\ref{convolqsat},\ref{ccc1},\ref{ccc2}) for an initial sand pile
making a finite angle with the firm soil shows that the profile of its
upwind left edge quickly evolves to get a horizontal slope which
smoothly continue the soil.  $h(x,t)$ and $x$ are in units of
$l_{sat}$.  The displayed profiles corresponds from left to right to
the times $t=0$, $0.25$, $0.5$, $0.75$ and $1.5$ in units of
$l_{sat}^2/Q_{sat}$.}
\label{ToePied}
\ec
\efig

An alternative is to consider the surface profile -- the firm soil as
well as the dune -- as a whole.  This is precisely what is done when
the convolution (\ref{convolqsat}) is used. If the slope $h'$ was not
continuous at the dune edge, in some sense, $h''$ would be infinite.
Then, from equation (\ref{ccc3}), the saturated flux would formally tend to
$-\infty$ over a region around this edge, i.e. would be forced to be null
because it cannot be negative. Thus in this region, only sand deposition is
permitted. This means that a discontinuity in the slope at the dune edge,
immediately reacts to prevent the motion of this boundary point, which starts
moving again only when the profile becomes flat, as shown on figure
\ref{ToePied}. Therefore, considering that the wind and thus the saturated
flux are not sensitive to the transition from the firm soil to the dune back,
gives a `natural' upwind boundary condition for propagative solutions: the
profile $h$ as well as its slope $h'$ should vanish at the upwind
boundary of the dune. This argument actually also applies at the downwind
edge, and we shall use it in the dome and dune next sections.

In the sequel, we are going to look for propagative solutions, i.e.
functions of the type $h(x-ct)$ and $q(x-ct)$.  We shall see that the
explicit form of these functions can be found for given upwind
conditions.  However, such solutions are parametrized by
coefficients such as this length $D$ or the propagative velocity $c$
which must be fixed by the right -- downwind -- conditions.  Two kinds
of right boundary conditions will be considered, leading to so-called
`dome dunes', i.e.  without avalanche slip face, or `actual' ones for
which a `recirculation bubble' will be introduced.

\subsection{General form of the propagative solutions}
For functions of $x-ct$, the continuity equation (\ref{ccc1}) can be
easily integrated and gives
\be
\label{qeth}
q=q_0 + ch.
\ee
We now describe everything in the propagating frame referential, and
rename $x$ as the new space coordinate.  We look for isolated
propagative objects.  The point $x=0$ is the beginning of the dune
where $h=0$.  $q_0$ is thus the incoming sand flux -- the sand supply.
In the region $x<0$, no grains are available on the ground: $h=0$ and
$q=q_0$ everywhere.  In the region $x>0$, using the relation between
$q$ and $h$ (\ref{qeth}) in the charge equation (\ref{ccc2}) with
$q_{sat}$ given by its expression (\ref{ccc3}), we get an ordinary
differential equation for the dune profile $h$:
\be
\label{equadiff}
1 - \alpha D h'' + (\beta-c) h' - ch - q_0 = 0,
\ee
where $h'$ and $h''$ denote the first and second derivatives of the
dune profile.

\bfig[t!]
\bc
\epsfxsize=\linewidth
\epsfbox{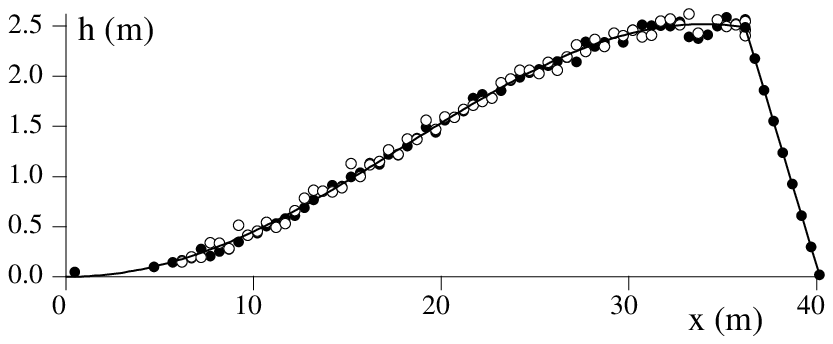}
\caption{Longitudinal profile of a barchan dune at Negrita beach,
southern Morocco. Black and white dots correspond to independent
measurements of the same dune. The solid line correspond to the best
fit by a function of the form (\ref{sincosexp}).}
\label{ProfilTerrain}
\ec
\efig

Under the two conditions $h(0)=0$ and $h'(0)=0$, the solution of equation
(\ref{equadiff}) is
\be
\label{sincosexp}
h(x) = \frac{1-q_0}{c} \left [ 1 + \left (
       \frac{s}{k} \sin(kx) - \cos(kx) \right ) e^{sx} \right ],
\ee
where the coefficients $s$ and $k$ are given by
\bea
\label{defs}
s & = & \frac{\beta-c}{2\alpha D}, \\
\label{defk}
k & = & \frac{1}{2\alpha D} \, \sqrt{4c\alpha D - (\beta-c)^2}.
\eea

Figure \ref{ProfilTerrain} shows the comparison between the central
longitudinal profile of an actual barchan and the theoretical form
(\ref{sincosexp}) which is parametrised by $k$, $s$ and $c$.  It can be seen
that the shape of the dune as well as the boundary conditions are well
captured by the model. In fact, an oscillatory function as (\ref{sincosexp})
does not look like an isolated dune yet. In particular, this $h(x)$ takes
negative values. Besides, as already mentioned, the two coefficients $D$ and
$c$ are not determined yet.  In the next sections we show how it is possible
to cut this solution at some point with an adequate right boundary
condition in order to get a genuine propagative profile.

\section{Domes}
\label{domes}
In this section, we look for so-called `dome' solutions, i.e. dunes which do
not show any avalanche plane. This is possible if the local slope of the dome
is everywhere not steeper than some threshold $\mu_b$. In the next section,
it will become clearer that this threshold corresponds to the slope at which
the wind stream lines detach the dune profile, creating a backward wind flow
-- or a `recirculation bubble' -- which leads to a slip face. As will be
explained in the conclusion, these dome solutions may play an
important role when extending the present model to three dimensional
situations. One of the major results of this section is that the shape of
the dome, when it exists, is selected by the value of the incident
flux $q_0$. Because we look for purely propagative solutions, $q_0$ of
course is also the flux of the sand which leaves the dune.

\bfig[t!]
\bc
\epsfxsize=\linewidth
\epsfbox{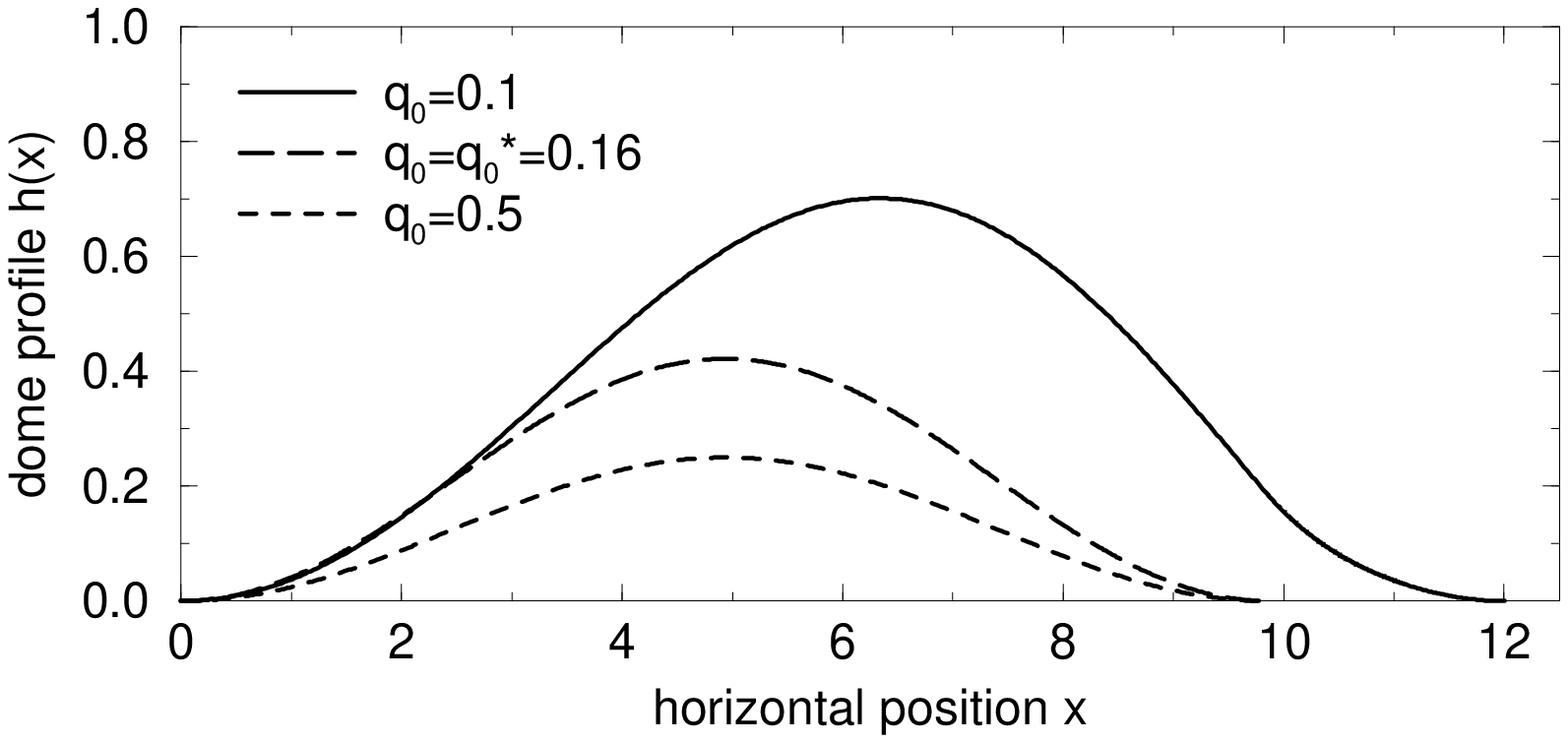}
\caption{Shape of the dome solution for different values of the sand flux
$q_0$. The data have been computed with $\alpha=1.$ and $\beta=4.$ Note that
these domes are actually very flat: for clarity the scale of the vertical
axis is much larger than that of the horizontal one.}
\label{domeshape}
\ec
\efig

Let us first take a very simple example. For $c=\beta$ the coefficient $s$
defined in equation (\ref{defs}) vanishes, and the solution (\ref{sincosexp})
reduces to $h(x)=\left [ 1 - \cos(kx) \right ] (1-q_0)/\beta$, where
$k=[\beta/(\alpha D)]^{1/2}$. As the general solution (\ref{sincosexp}) was
constructed for, $h$ and $h'$ vanish at $x=0$, but also at $x=2\pi/k$. This
value sets the length of the dune: $D=4\pi^2\alpha/\beta$. Interestingly,
this is precisely the cut-off $\lambda_c$ of the linear stability analysis.
As discussed in the previous section, this simple solution satisfies the
`natural' downwind boundary conditions $h(D)=0$ and $h'(D)=0$. Another
important restriction is that on the saturated flux $q_{sat}$ which must
remain positive everywhere. As a matter of fact, there is a minimum value
of $q_0$ below which $q_{sat}$ takes negative values and makes this solution
inconsistent. It is easy to show that this lower bound for $q_0$ reads
\be
\label{q0star}
q_0^* = 1 - \frac{1}{\sqrt{ 1 + \left ( \frac{\beta}{2\pi\alpha}
\right )^2}}.
\ee
For the values $\alpha=1.$ and $\beta=4.$, we get $q_0^* \simeq 0.16$. Two
such dome profiles for $q_0=0.5$ and $q_0=q_0^*$ are displayed on figure
\ref{domeshape}. At last, note that the steepest slope of this solution is
$h'_{min}=-\frac{1-q_0}{2\pi\alpha}$. For $q_0=q_0^*$ and the same
numerical values as above, it gives $|h'^*_{min}| \simeq 0.13$. For any
slope threshold $\mu_b$ larger than $|h'^*_{min}|$, the whole set of
these solutions such that $q_0^* \le q_0 \le 1$ is acceptable. For $\mu_b$
below this (actually rather small) value however, $q_0$ must be larger than
$q_0^c=1.-2\pi\alpha\mu_b$, and no fully consistent dome solution can be
constructed for a smaller incident sand flux. In the sequel, the typical
value of $\mu_b$ that we shall use will be $\mu_b=0.25$.

\bfig[t!]
\bc
\epsfxsize=\linewidth
\epsfbox{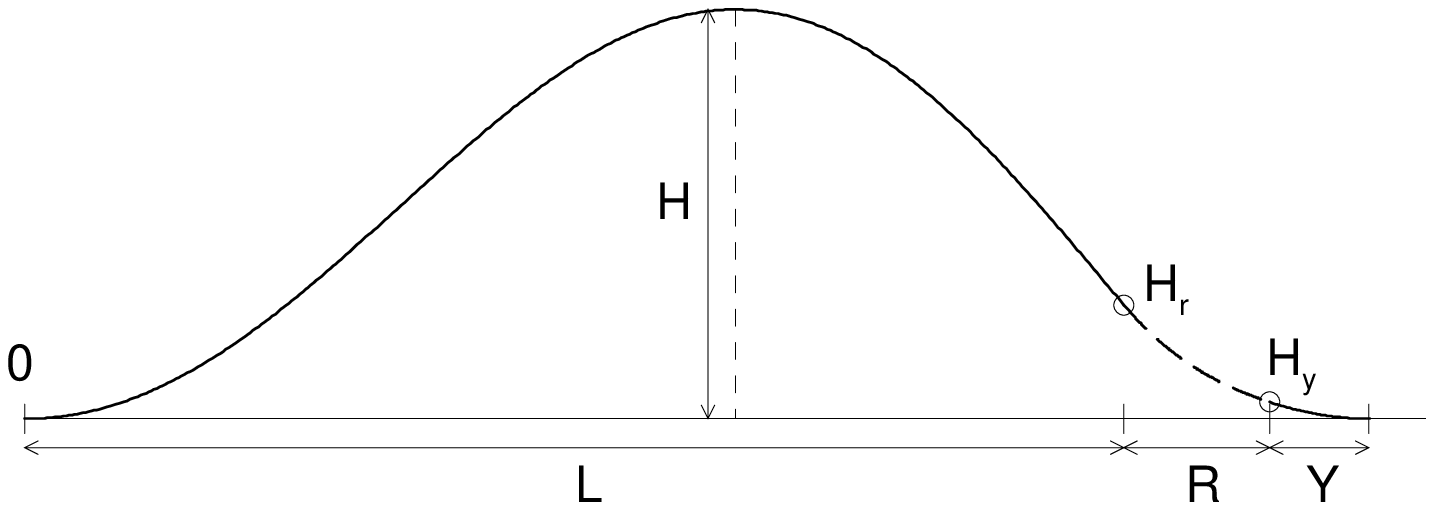}
\caption{Definition of the parameters of the dome. $x=0$ is the reference
point of the dune. $H$ is its height. For $q_0 < q_0^*$, its length $D$ must
be cut into three regions of size $L$, $R$ and $Y$. As explained in the text,
in the middle region (dahed line) the saturated flux vanishes and the profile
is a branch of exponential which matches with the two parts of the full
solution. $H_r$ and $H_y$ are the heights at the two sides of this central
region.}
\label{domeparam}
\ec
\efig

How can we construct a dome solution for $q_0 < q_0^*$? Suppose the values of
the velocity $c$ and the length $D$ are given, from equation
(\ref{sincosexp}) we can compute $h$ and its derivatives as well as $q_{sat}$.
Because $q_0$ is smaller than $q_0^*$, $q_{sat}$ will reach zero at some
position $x=L<D$. Negative values of $q_{sat}$ are not permitted, and
we therefore set it to zero for $x \ge L$. Then, equation (\ref{ccc2}) is
very easy to integrate and gives and exponential branch for the flux $q(x)$,
which, using the linear relation (\ref{qeth}) between $q$ and $h$ leads to:
\be
\label{brancheR}
h_r(x) = \frac{1}{c} \left [ (cH_r+q_0) e^{-(x-L)}-q_0 \right ].
\ee
The subscript $r$ is used to avoid any confusion with $h(x)$ given by
equation (\ref{sincosexp}), but both are part of the same dome solution.
In particular, the dome must be smooth and in this expression $H_r=h(L)$
-- we shall come back in the next paragraph on the continuity conditions
at $x=L$. We cannot end the dome solution with $h_r$ for two reasons. First,
we argued in the previous section that the dome profile must ends with a
horizontal slope, which is not possible with an exponential function at
finite distance $D$. Second, if one compute what would be the saturated
flux $q_{sat}^r$ calculated from the profile $h_r$ with a relation like
(\ref{ccc3}), one sees that it is negative as it should close to $x=L$
(remember that $q_{sat}$ is set to zero), but crosses zero at some other
position $x=L+R<D$. From that point and for larger $x$ we thus need to
come back to the original profile $h$. Then, a natural way to end up the
dome profile is to use
\be
\label{brancheY}
h_y(x) = h(D-x),
\ee
which is consistent if $q_{sat}(-Y)=0$, with $L+R+Y=D$. This choice ensures
that both $h(D)$ and $h'(D)$ vanish as required. These three regions of the
dome solution are illustrated on figure \ref{domeparam}.

\bfig[t!]
\bc
\epsfxsize=\linewidth
\epsfbox{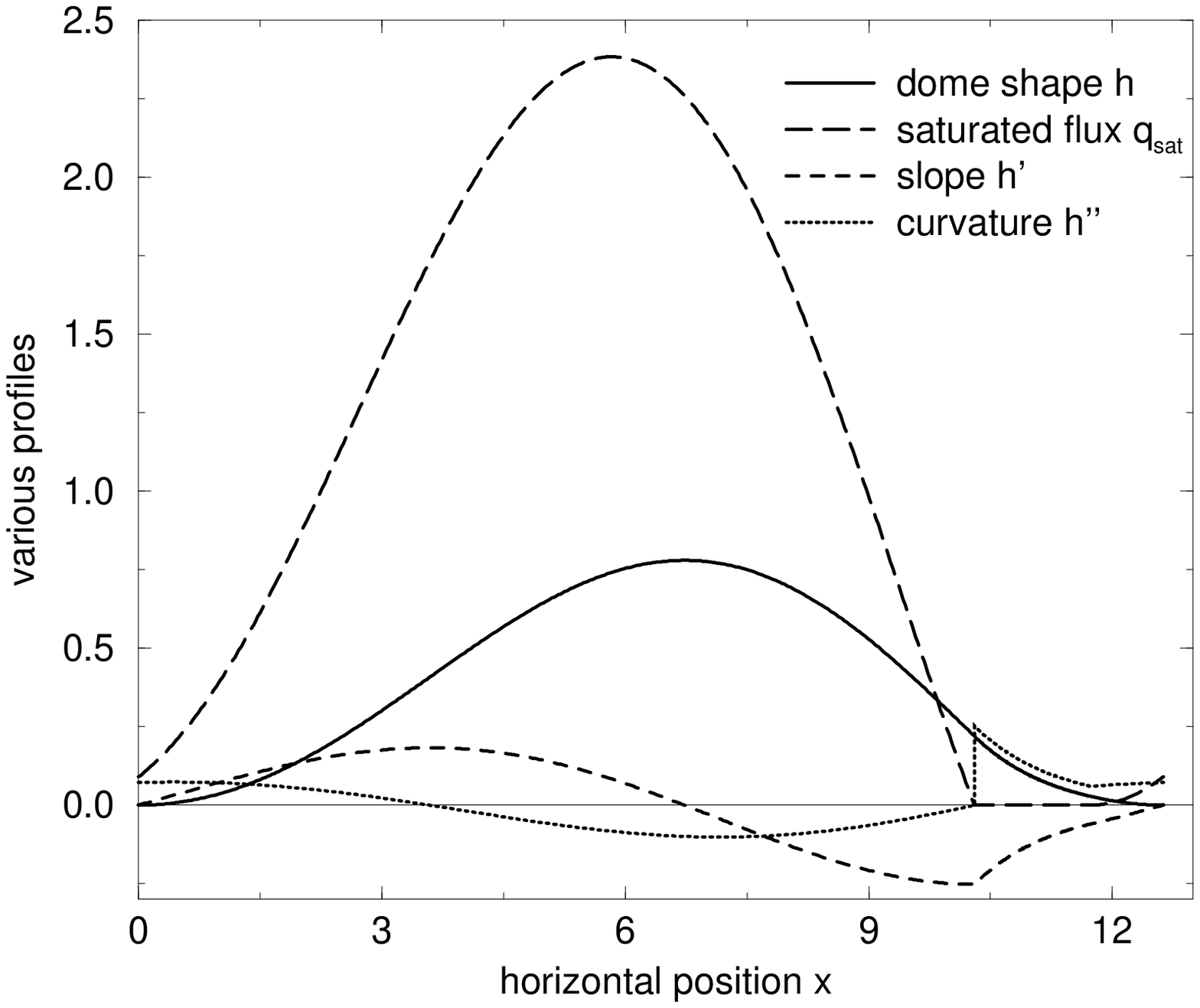}
\caption{Dome profile and its derivatives, as well as the saturated sand flux
for $q_0=0.09$. With this value of the incident sand flux, the dome steepest
slope just reaches its minimum permitted value $-\mu_b$ at $x=L$. $q_{sat}$
is strictly zero between $x=L$ and $x=L+R$ and $h$ is a branch of exponential.
At $x=L$ the curvature $h''$ shows a discontinuity, but all quantities are
continuous at $x=L+R$.}
\label{domeprofiles}
\ec
\efig

Let us now be more explicit about the continuity conditions at the two
matching points. At $x=L$, by construction of the relation (\ref{brancheR}),
the dome profile, the sand flux, and the saturated flux are continuous.
Because equation (\ref{ccc2}) holds everywhere, it implies that $\dr_x q$,
and therefore the slope $h'$ are also continuous. By contrast, the curvatures
$h''(L)$ and $h''_r(L)$ are different, and therefore
$q_{sat}^r(L) \ne q_{sat}(L) = 0$. The position $x=L+R$ is defined by
$q_{sat}^y(L+R)=q_{sat}(-Y)=0$. At this point of course we do not want any
step in the dome profile, such that $H_y \equiv h_y(L+R)$ and $h_r(L+R)$ must
be equal. Again, because of equation (\ref{ccc2}) and the fact that $q_{sat}$
has been built to be continuous at $x=L+R$, the continuity of the profile
makes the slope continuous too. However this position is also the point where
the pseudo saturated flux $q_{sat}^r$ crosses zero, such that the curvature
of the dome profile is also continuous. All these continuity conditions can
be shown on figure \ref{domeprofiles}.
\bfig[t!]
\bc
\epsfxsize=\linewidth
\epsfbox{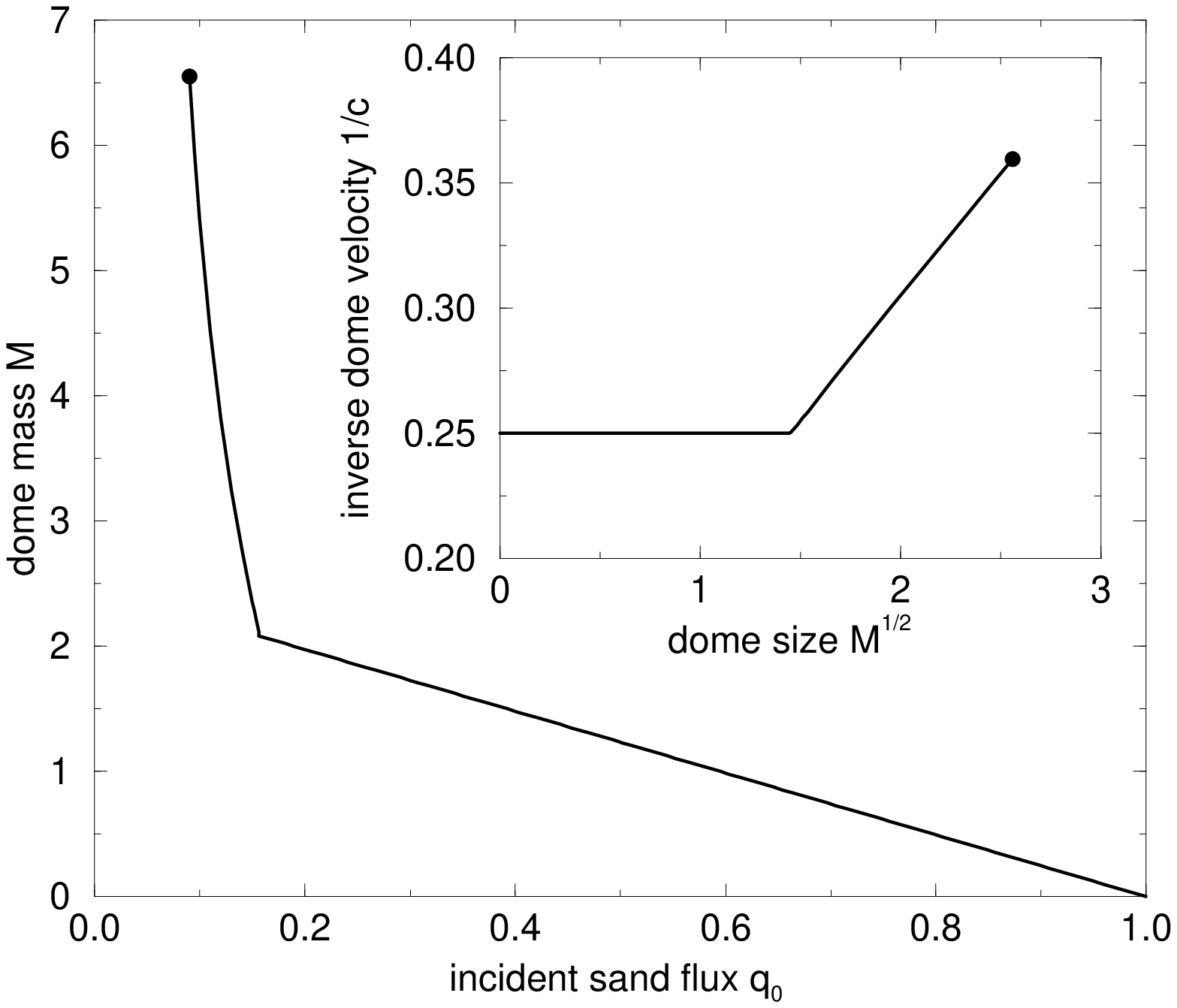}
\caption{The main graph shows the mass of a dome selected by an incident
flux $q_0$. This function is a straight line for $q_0 \ge q_0^*$ which
corresponds to the $c=\beta$ solutions. The solutions for $q_0 \le q_0^*$
are cut off at $q_0=q_0^c$ where the slope at $x=L$ is equal to the threshold
$-\mu_b$ (big dot). The important point is that this curve $M(q_0)$ is a
decreasing function, which make these domes unstable to a change of $q_0$.
In the inset, the inverse of the velocity $1/c$ is plotted against a typical
size $M^{1/2}$ of the dome. It is strictly constant for $M \le M(q_0^*)$ and
then almost straight up to the cut-off value $M_c$.}
\label{domeMandc}
\ec
\efig
\bfig[t!]
\bc
\epsfxsize=\linewidth
\epsfbox{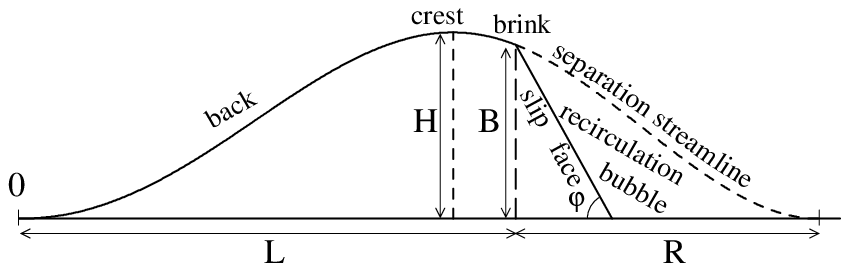}
\caption{Definition of the dune and recirculation bubble parameters.
$x=0$ is the reference point of the dune.  The brink -- where the slip
face and the recirculation bubble start -- is located at $x=L$.  The
separaration streamline extends the dune profile, starting with the
same slope $p$ and curvature $C$.  It smoothly reaches the ground at
$x=L+R$.  We note $H$ the height of the highest point of the dune
(crest).  $B$ is the dune height at the brink.  $H$ and $B$ may
coincide or not.  We call $\phi$ the angle of the slip face. For the
numerics, we use $\phi=30^o$.}
\label{duneparam}
\ec
\efig

To sum up, we have in practice four coefficients to determine: $c$, $L$, $R$
and $Y$ ($D=L+R+Y$) with the four non-linear following equations:
\bea
\label{eqdome1}
q_{sat}(L)     & = & 0, \\
\label{eqdome2}
q_{sat}(-Y)    & = & 0, \\
\label{eqdome3}
h_r(L+R)       & = & h(-Y), \\
\label{eqdome4}
q_{sat}^r(L+R) & = & 0,
\eea
which can be done numerically. One can find such a solution for any
$q_0 < q_0^*$. However, to be consistent with the next section, and as
already mentioned in the $c=\beta$ case, such a dome is acceptable
only if its steepest slope is larger than $-\mu_b$. This fixes a lower
bound $q_0^c$ under which there are no solution. With $\alpha=1.$,
$\beta=4.$ and $\mu_b=0.25$, we find that $q_0^c \simeq 0.09$. The solution
for this particular value of the incident sand flux is plotted on figure
\ref{domeprofiles}, and the dome profile for $q_0=0.1$ is compared, on figure
\ref{domeshape}, with two cases of the $c=\beta$ solutions.

When we have all coefficients of a particular solution, we can compute the
mass of the corresponding dome:
\be
\label{massedome}
M = \int_0^D \!\!\! dx \,  h(x).
\ee
For the $c=\beta$ solutions, this mass can be simply expressed as
$M=4\pi^2\alpha(1-q_0)/\beta^2$.  The interesting point is that the
function $M(q_0)$ decreases with its argument, see figure
\ref{domeMandc}.  Then, take a propagating dome under a flux $q_0$,
and suppose that, for some reasons, the incident flux is suddenly
increased by some small amount $\delta q_0$.  The shape of the profile
is such that the flux of the sand which leaves the dome is exactly
$q_0$.  Then, its mass gets larger.  But a larger mass means that the
flux of sand which can leave the dome is even smaller than before,
which makes the mass of the dome increase again, and so on.  In the
same way, if the initial flux $q_0$ is suddenly decreased, the dome
will loose more and more sand because less mass means a higher leaving
sand flux.  This negative feedback mechanism suggests that domes are
unstable objects that may either quickly disappear or reach the point
where their slope is steep enough to generate a recirculation bubble
and create an avalanche slip face to become an actual dune.  One could
however observe them, for example under water with periodic boundary
conditions \cite{BKFR01}, which can stabilize such an instability.
Then an interesting prediction on the velocity of these domes which
could be compared to experiments is shown in the inset of figure
\ref{domeMandc}: it is constant for small domes and $1/c$ grows almost
linearly with their size $M^{1/2}$ for larger ones.  The dome length
$D$ behaves also very much like $1/c$ as a function of $M^{1/2}$.

\section{Dunes}
\label{dunes}
It is possible to get another type of propagative profiles --
transverse dunes -- with a quite more sophisticated downwind boundary
condition, namely a recirculation bubble.  It is a common field
observation (see part 1) that the wind streamlines on a dune follow
exactly the shape of its back profile but separate at the point where
the avalanche slip face begins -- the brink -- and reattach further
downwind, as shown on figure \ref{duneparam}.  This phenomenon creates
an eddy in the `shadow' of the dune, where the wind is much less
strong than anywhere else.  As a consequence, all the sand eroded on
the back is deposited around the top of the slip face which avalanches
when the slope becomes too steep.  As explained in the first part of
these twin papers, it is fortunate that an accurate description of
these avalanches is not necessary due to the fact that they do not
have any feedback on the back profile of the dune: they simply relax
their equilibrium slope $\tan\phi$.  Because no grains can escape the
dune from the slip face through the recirculation bubble, the net out
flux is zero, which fixes $q_0=0$.  Note that this is particular to
two-dimensional situations that we are focused on: three dimensional
barchan dunes loose sand from their horns.

\subsection{The recirculation bubble}
The simplest way to model the feedback of the recirculation bubble on
the whole dune has been proposed by Zeman and Jensen \cite{JZ85} and
used more recently by Kroy \textit{et al}.  \cite{KSH01}.  The idea is
to build an envelope of the dune which prolongs the dune profile by the
separation streamline -- see figure \ref{duneparam}.  To the first order,
the wind on the back of the dune is the same as that would have been
obtained if the envelope was solid.  For example, the convolution integral
in equation (\ref{convolu2}) used to calculate the shear stress on the
soil should be applied not to the dune profile but to the dune + bubble
envelope.

In our approach, the effect of the recirculation bubble on the wind is
simply to modify the total length of the dune $D$ which becomes the dune
length $L$ plus the bubble length $R$: $D=L+R$.  The importance
of the recirculation bubble becomes very clear: because it makes
the apparent length of the dune be larger, it increases the erosion
at the top of the dune, due to the curvature effect.

We are going to construct a very simple description of the
recirculation streamline.  Because it extends smoothly the dune shape,
it will be possible to link its parameters to the dune ones, and to
impose this way what could be called `non-local right boundary
conditions' to the solution expressed by equation (\ref{sincosexp}).

Let us call $h_b(x)$ the separation streamline profile. As shown on figure
\ref{duneparam}, $x=L$ is the point at which this line starts and $x=L+R$
is its reattachment point with the soil.  At both matching points,
the extension of the dune or the soil by the separation streamline should be
smooth:
\bea
\label{bubbleenL1}
h_b(L)    & = & h(L)  \equiv B, \\
\label{bubbleenL2}
h_b'(L)   & = & h'(L) \equiv p, \\
\label{bubbleenL+R1}
h_b(L+R)  & = & 0, \\
\label{bubbleenL+R2}
h_b'(L+R) & = & 0.
\eea

The recirculation bubble will be essentially governed by one important
parameter: a slope that we already named $\mu_b$ in the previous section.
It corresponds to the maximum angle that the wind streamlines can follow
behind an obstacle. In other words, there is a flow separation if the slope
is locally steeper than $-\mu_b$.  It is natural to impose that the steepest
slope of the separation streamline should then be $-\mu_b$:
\be
\label{bubbleinflexion}
h_b'(x_b)=-\mu_b \quad \mbox{with} \quad h_b''(x_b)=0.
\ee
As a matter of fact, this steepest slope should be in principle less than
or equal to $-\mu_b$ otherwise this would mean that the streamlines actually
detach at a higher threshold value. Besides, the bubble initially appears
exactly at this slope on critical domes. The relation (\ref{bubbleinflexion})
can be then taken as the simplest self consistent slope condition.

Because the wind velocity profile is scale invariant, a further property
is that the length $R$ of the recirculation bubble should scale on the
dune size and no new length scale should be introduced by the bubble.
Therefore, the curvature at $x=L$ must also be continuous:
\be
\label{bubbleenL3}
h_b''(L) = h''(L) \equiv C.
\ee
Such a condition is also naturally observed when numerically integrating
equations (\ref{convolqsat},\ref{ccc1},\ref{ccc2}) -- i.e. with the
convolution term of equation (\ref{convolqsat}) rather than the simplified
relation (\ref{ccc3}). Furthermore, we noticed that the stability of the
numerical scheme is very sensitive to the way the slope is computed at $x=L$:
the only choice which does not create any instability is that which uses a
formula which mixes the dune and bubble profiles
$h'(L) = [h_b(L+\Delta x) - h(L)]/\Delta x$, where $\Delta x$ is the
discretisation step.

In the absence of available data or models on flow separation behind a
dune, the explicit form of the bubble profile we chose is the simplest
that satisfies the requirements written above, namely a polynomial of
$3^{\mbox{\scriptsize rd}}$ degree:
\be
\label{bubble}
h_b = B + p (x-L) + \frac{1}{2} C (x-L)^2 + \frac{1}{6} G (x-L)^3.
\ee
Such a choice was also that of Kroy \textit{et al}. in \cite{KSH01}. Using
the conditions (\ref{bubbleenL1}-\ref{bubbleenL3}), $G$, $C$ and $R$ can
be computed as functions of the slope $p$ and the height $B$ at the brink:
\bea
\label{defG}
G & = & \frac{9\mu_b}{R^2} \left ( 1 - \frac{2B}{3R\mu_b} \, \sqrt{1 -
\frac{4B}{3R\mu_b}} \right ) \\
\label{defC}
C & = & -RG + \sqrt{2G\mu_b} \\
\label{defp}
p & = & -\mu_b + \frac{C^2}{2G}
\eea
We have checked that other similar choices for the parame\-trization
of the separation streamline do not change the qualitative conclusions
that are going to be presented in the next subsections.  In a general
way, the recirculation bubble conditions can be expressed as $R/B =
f_1 (p)$ and $C B = f_2 (p)$, where the functions $f_1$ and $f_2$
encode the particular choice of the streamline separation profile
$h_b$.

These two relations put together with the explicit expression of the dune
profile (\ref{sincosexp}) and the differential equation that it verifies
(\ref{equadiff}), where $D$ has been set to the whole dune + bubble length
$D=L+R$, let one to get three relations which link together the four parameters
$c$, $L$, $R$ and $B$. Instead of plotting, say, the three first with respect
to the last one, we rather chose to express, as we did for the domes in the
previous section, all of them as functions of the total mass $M$ of the dune:
\be
\label{massedune}
M = \int_0^L \!\!\! dx \,  h(x) + \frac{B^2}{2\tan\phi}.
\ee
We then get a continuous set of dune solution, from very large values
of the mass, down to some cut-off value $M_c$ below which no stable
recirculation bubble can be constructed -- see below.

\subsection{Profiles, dimensions and velocities}
\bfig[t!]  \bc \epsfxsize=\linewidth \epsfbox{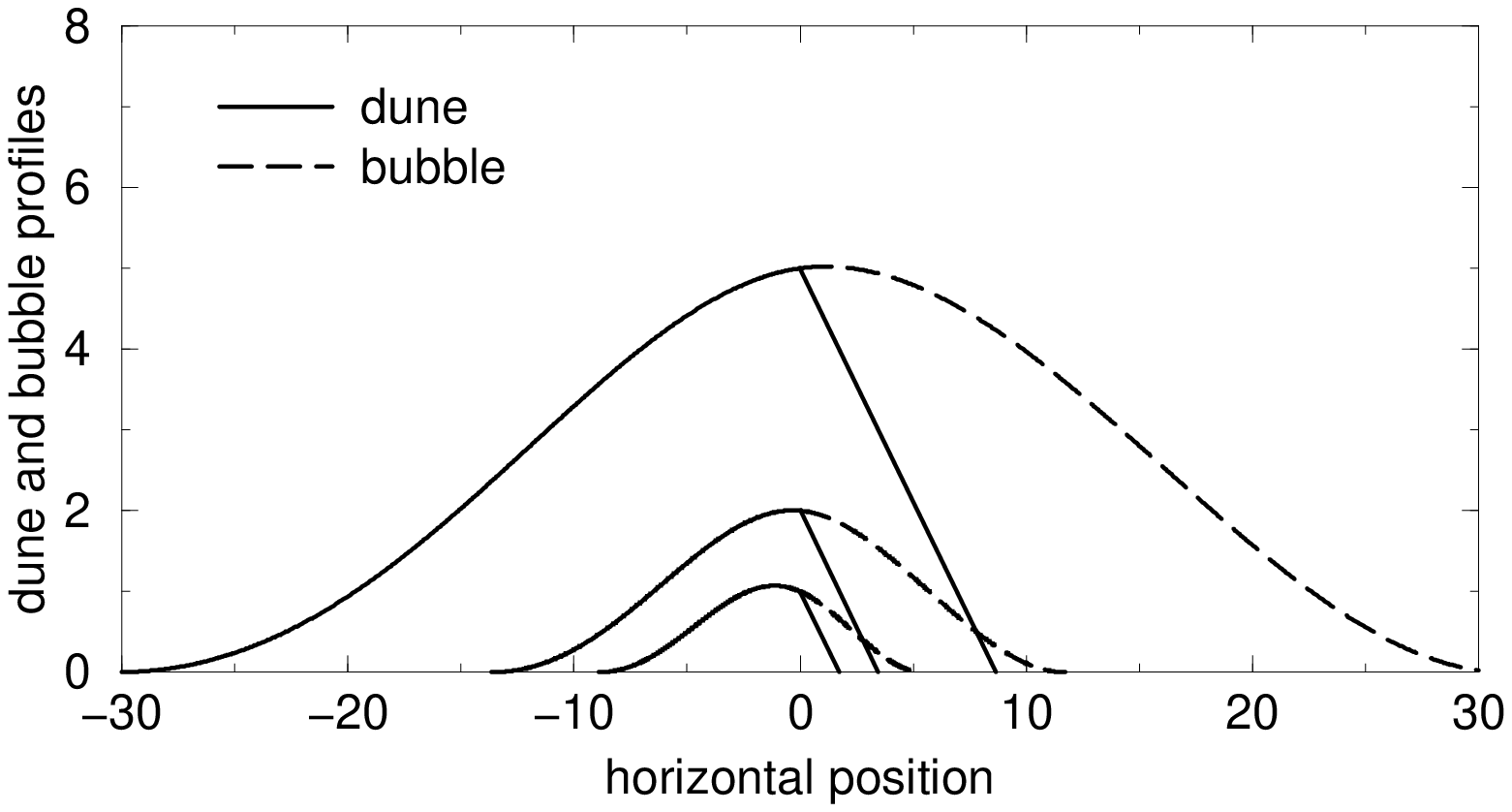}
\caption{Dunes of different masses and corresponding recirculation
bubbles.  From small to large, the masses are $M=6.05$, $16.6$ and
$88.7$.  The other parameters are $\alpha=1.$, $\beta=4.$ and
$\mu_b=0.25$.  For clarity, all these profiles have been shifted in
order to get all brink positions at $x=0$.
\label{profils.brut}}
\ec
\efig
\bfig[t!]
\bc
\epsfxsize=\linewidth
\epsfbox{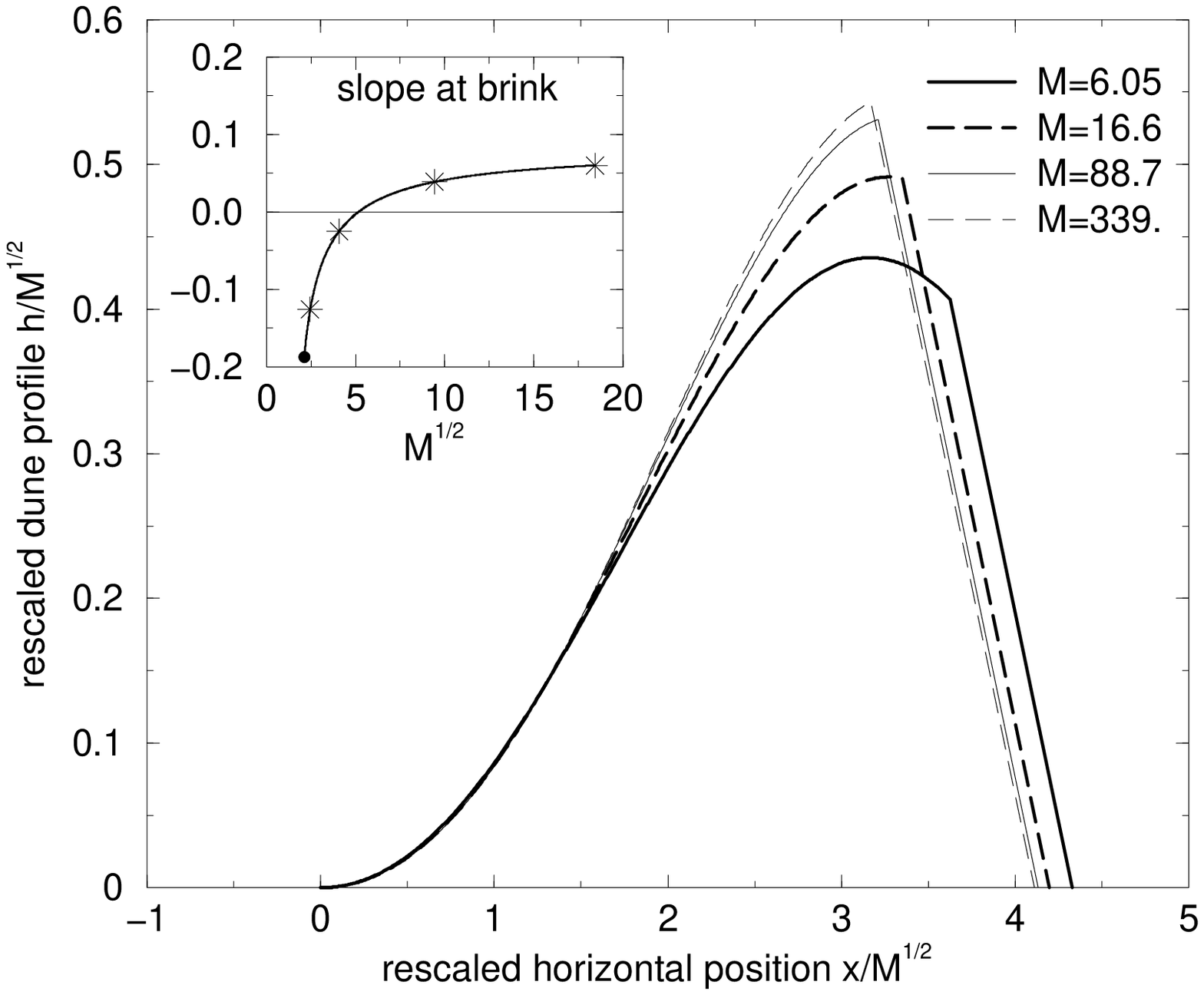}
\caption{Rescaled profiles of dunes of different sizes. The data have been
computed with $\alpha=1.$, $\beta=4.$ and $\mu_b=0.25$. Lengths have been
rescaled by the square roots of the masses of the dunes. These profiles are
not scale invariant: as shown in the inset, the slope $p$ just before the
brink is negative for small dunes and positive for large ones, such that
depending on the dune size, the crest does or does not coincide with the
brink. The four stars represent the four dune profiles.
\label{profils.resc}}
\ec
\efig
\bfig[t!]  \bc \epsfxsize=\linewidth \epsfbox{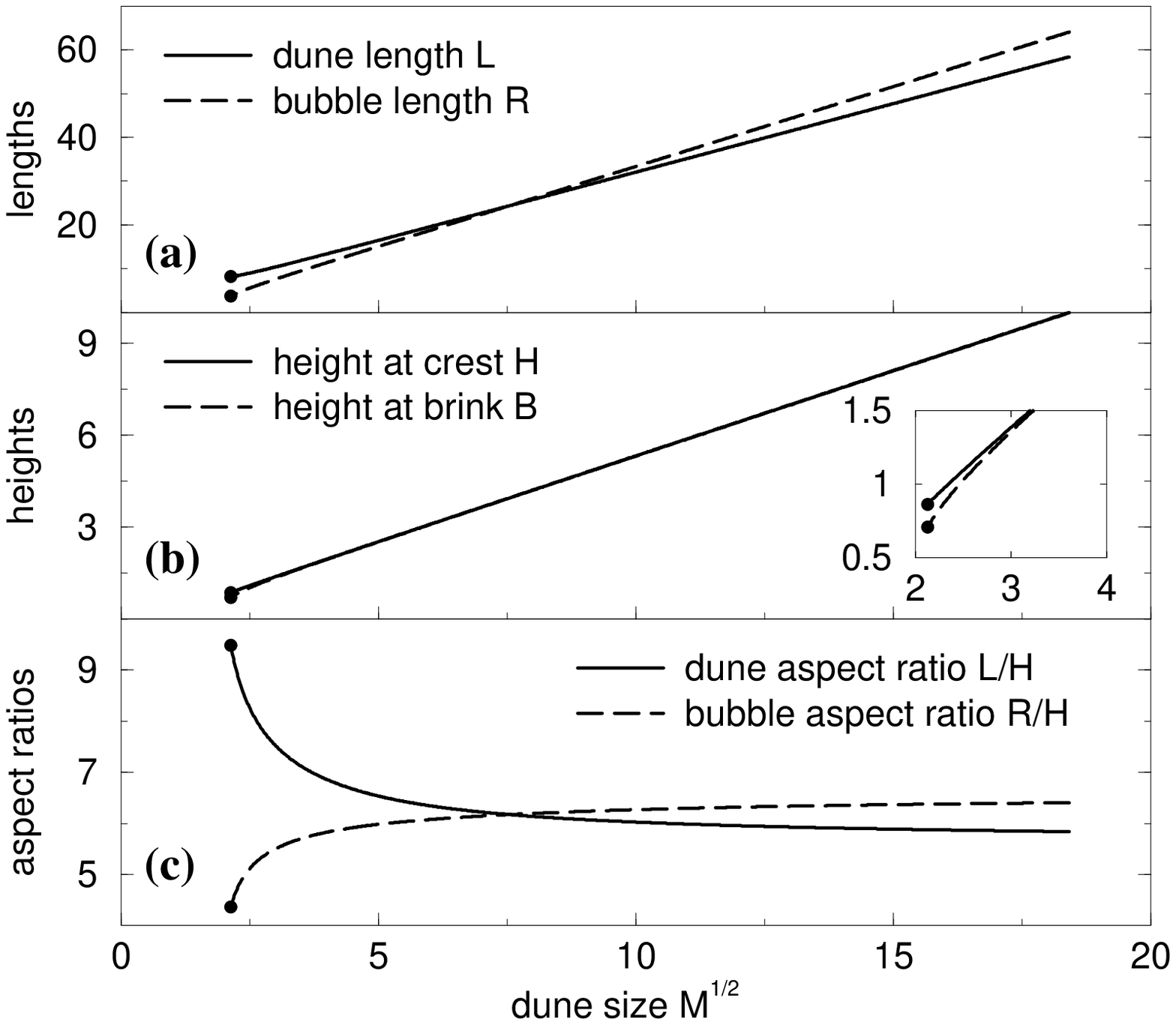}
\caption{Scaling of the lengths $L$, $R$ (a) and $H$, $B$ (b) with the
dune size $M^{1/2}$.  As a first approximation, these plots are almost
straight lines.  No recirculation bubble -- and therefore no dune
solution -- can be constructed below some cut-off value (big dot).
The inset on graph (b) is a zoom around this cut-off scale.  Graph (c)
shows that the dune and bubble aspect ratios are not constant: large
dunes are more compact with a proportionally larger bubble than small
ones.  The data have been computed with $\alpha=1.$, $\beta=4.$ and
$\mu_b=0.25$.
\label{dimensions}}
\ec
\efig
\bfig[t!]
\bc
\epsfxsize=\linewidth
\epsfbox{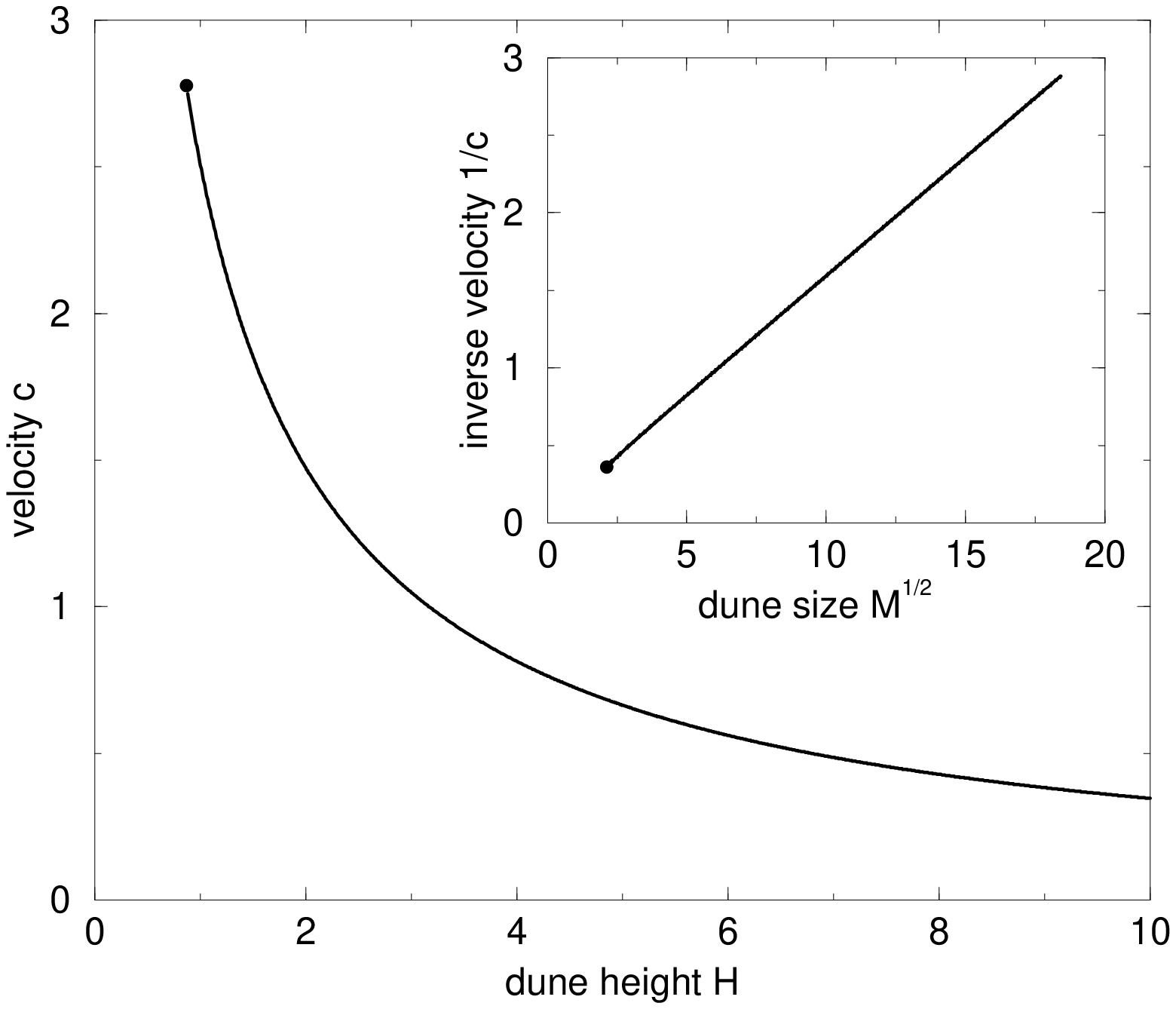}
\caption{Propagation velocity $c$ of the dune as a function of its height $H$.
As shown in the inset, with a very good precision, $1/c \sim aM^{1/2}+b$
down to some cut-off value. These data have been computed with
$\alpha=1.$, $\beta=4.$ and $\mu_b=0.25$.
\label{vitesses}}
\ec
\efig

Let us fix a value of the mass of sand $M$ available to construct a
dune.  As for the domes, even for the simple choice of the
recirculation bubble profile (\ref{bubble}), a numerical resolution is
required at this point to get the corresponding values of the
parameters $c$, $L$, $R$ and $B$.  Three examples of such solutions
are plotted on figure \ref{profils.brut}.  As evidenced on figure
\ref{profils.resc}, when rescaled by a typical dimension of the dune
-- here the square root of the mass --, the different profiles do not
collapse on a single curve.  In particular the slope at the brink
$p=h'(L)$ varies with respect to $M$, and even changes its sign for
this choice of the parameters $\alpha$, $\beta$ and $\mu_b$.  We get
$p<0$ for the smallest dune, while $p>0$ for the largest one.

The inset of figure \ref{profils.resc}, as well as the curves of
figures \ref{dimensions} and \ref{vitesses}, which all show the dune
and bubble features (lengths, slope, aspect ratios and velocity) as a
function of the mass of the dune $M$, are all cut off at some value
$M_c$ (big dot).  This point corresponds to the smallest solution that
can be constructed.  Technically, one can see that the equations
(\ref{defG}-\ref{defp}) which link bubble and dune parameters loose
their sense when the quantity $3R\mu_b - 2B$ becomes negative.  This
point precisely gives the cut-off mass under which no bubble, and
therefore no dune solution can be constructed.  It is important to
notice that this smallest dune has an avalanche slip face of finite
size.  Another choice for the function $h_b(x)$ would of course have
given different values for $M_c$ as a function of the parameters of
the model.  But physically, this cut-off scale corresponds to the fact
that such a recirculation bubble must have a minimum spatial extension
to accommodate all continuity constrains.  Not surprisingly, this
extension is of the order of unity, that is to say of the order of
$l_{sat}$ in non rescaled length units -- which is the only length
scale of our system.  At last, it must be mentioned that the
parameters of this smallest dune are pretty close to that of the
largest dome, i.e.  the dome for which the steepest slope is precisely
equal to $-\mu_b$, see table \ref{domedune}.

\begin{table}
\bc
\btab{|l|c|c|c|}
\hline
              & M       & c       & H       \\
\hline
Smallest dune & $ 4.51$ & $ 2.78$ & $ 0.86$ \\
\hline
Largest dome  & $ 6.55$ & $ 2.78$ & $ 0.78$ \\
\hline
\etab
\caption{Mass, velocity and height of the smallest dune solution that one can
construct, compared to that of the largest dome for which the steepest slope
is equal to $-\mu_b$. Consistently, these data are pretty close to each
other. These numerical values have been computed with $\alpha=1.$, $\beta=4.$
and $\mu_b=0.25$.
\label{domedune}}
\ec
\end{table}

All results of figures \ref{dimensions} and \ref{vitesses} are remarkably
simple and resemble very much field observations: the lengths $L$, $R$, $H$
and $B$ are almost straight lines as a function of the dune size $M^{1/2}$.
Similarly, $1/c \sim aM^{1/2}+b$ with a very good precision. At last, it must
be noted that the dune and bubble aspect ratios are not constant: large dunes
are more compact with a proportionally larger bubble than small ones.

\subsection{Parametric study}

To complete the results of the previous subsection, we present here a
parametric study of few quantities, namely the minimal dune height $H_c$
and the dune slope at the brink of asymptotically large dunes ($M \to \infty$).
The parameter $\alpha$ is kept to unity, but $\beta$ has been varied between
$2$ and $6$, and $\mu_b$ from $0.2$ to $0.4$.

\bfig[t!]
\bc
\epsfxsize=\linewidth
\epsfbox{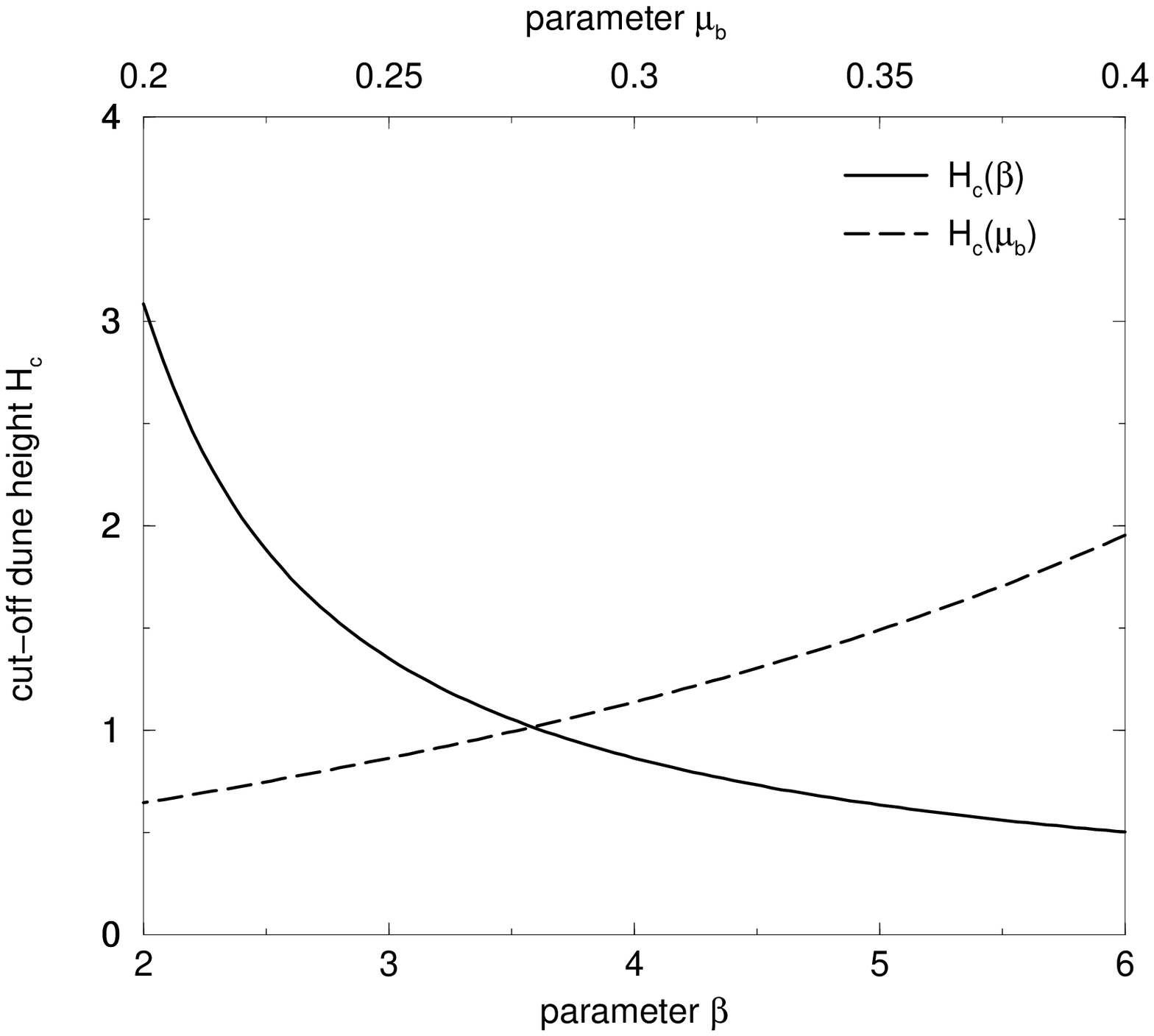}
\caption{Variation of the minimal dune height $H_c$ with the parameters $\beta$
and $\mu_b$. To plot theses curves, we took $\alpha=1.$ Besides, the solid
line has been obtained with $\mu_b=0.25$, and the dashed one with $\beta=4.$
\label{Hcdebetaetmub.a=1}}
\ec
\efig
\bfig[t!]
\bc
\epsfxsize=\linewidth
\epsfbox{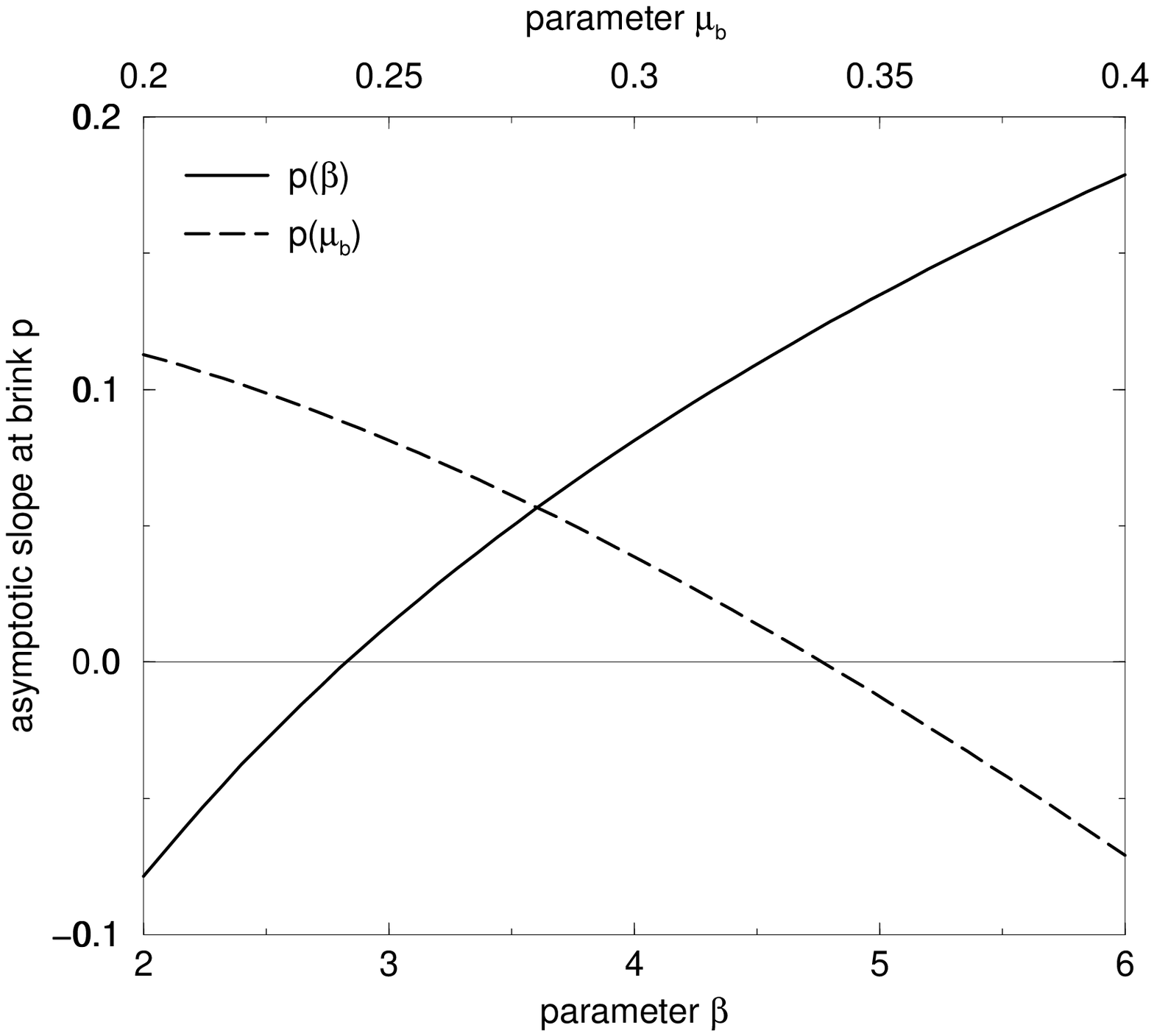}
\caption{Variation of the asymptotic ($M \to \infty$) slope at the brink $p$
with the parameters $\beta$ and $\mu_b$. To plot theses curves, we took
$\alpha=1.$ Besides, the solid line has been obtained with $\mu_b=0.25$, and
the dashed one with $\beta=4.$ Interestingly, $p$ remains negative for small
$\beta$ or large $\mu_b$.
\label{ppinfty.a=1}}
\ec
\efig

Figure \ref{Hcdebetaetmub.a=1} shows that the variation of $H_c$ as a function
of the parameters of the model is not very strong. Similarly, the aspect
ratio of very large dunes is always of the same order of magnitude.
The most interesting fact is perhaps that for small values of $\beta$, or for
large values of $\mu_b$, the slope at the brink of very large dunes can remain
negative -- see figure \ref{ppinfty.a=1}. In other words, the brink and the
crest can be always distinct, even for asymptotically large dunes.

One can understand intuitively the variations of $H_c$ and $p$ with $\beta$
and $\mu_b$. Increasing $\beta$ gives more strength to the destabilizing
process, which lets small dunes appear at lower critical scale (smaller $H_c$)
and makes large dunes more bumpy (larger $p$). If $\mu_b$ gets larger, the
situation is reversed.

\section{Conclusion}
\label{conclu}

We have shown in this paper how, inspired from the work of Sauermann
\textit{et al}., one can build a simpler two dimensional model for the
formation and the propagation of dunes. This modelling is based on two
main variables: the dune profile $h$ and the volumic sand flux $q$. It
includes three effects: (i) the mass conservation, (ii) the space lag over
a length $l_{sat}$ for the sand flux to become saturated at some value
$q_{sat}$, and (iii) the feedback of the profile on the saturated flux. In
this third phenomenological equation, erosion and deposition processes are
the result of the competition between two antagonist mechanisms: a
stabilizing non local curvature term ($\alpha$) and a destabilizing slope
one ($\beta$) which breaks the upwind-downwind symmetry.

Two kinds of solutions have been found: so-called `domes' which do not show
any avalanche slip face, and `real' dunes for which a downwind
recirculation bubble has been introduced. We were able to predict an
analytical form for their propagative profiles, but whose coefficients have
to be computed numerically.

The results of the model resemble very much field observations. We found for
example that, due to this length scale $l_{sat}$, the dune profiles are not
scale invariant: small dunes are flatter than large ones. Another point
is that the inverse of the propagative velocity $c$ is, to a very good
precision, almost linear with the size of the dune. This is consistent with
Bagnold's argument that $c \simeq q_{sat}/H$ for a dune of height $H$. In
fact, this relation overestimates the velocity of small dunes for which
the sand flux may not be already saturated at the crest and also because,
compared to large dunes, the value of $q_{sat}$ is reduced due to a smaller
curvature.

An important point discussed all through the paper was the issue of
the boundary conditions.  In particular, an important physical input
was the `recirculation bubble' behind the dune.  This bubble makes the
dune effectively look larger to the wind, and, due to the non local
term in the relation between the saturated flux and the dune profile,
has a stabilizing role.  In fact, very little is known and well
established about this recirculation bubble, but most of the dune
features (position of the slip face, cut-off size, etc.)  precisely
depend on fine interactions and feedbacks between the dune and the
bubble.  More studies on this point are then urgently needed.

Another central result of the paper is the existence of this cut-off scale
just mentioned, below which no dune can form. It corresponds to the fact
that the bubble must have a minimum spatial extension to accommodate all
continuity constrains. Not surprisingly, this scale is of order of $l_{sat}$.
This result then rises the question of dune initiation and formation.
The two scenari usually proposed by geophysicists for the formation of
dunes are the following: first possibility, a small bump (of the size
of ripples) grows continuously and forms a dune; second one, the sand
accumulates on a solid obstacle like a rock or a bush and, when the
size of the accumulation becomes larger than the obstacle, a dune
forms and starts propagating downwind.  However, observations show
that ripples are stable and no structures between dunes and ripples
can be seen.  Similarly, rocks and bushes creates lee dunes of the size
of the obstacle which remain anchored to the obstacle. An alternative
explanation can be proposed, following the results of the stability
analysis of the equations of the model, as well as that of the dome solutions.
We found that large wavelengths perturbations get amplified, and that the
dome profiles, selected by their incident flux $q_0$, are unstable to
changes of that flux. Then, a possibility is that first domes form with a
small height but directly with large length and width, and second that
these domes progressively become more and more compact, and eventually reach
the point where their slope is steep enough to generate a bubble and create
an avalanche slip face to become an actual dune.

Several extensions to the present work can be thought of.  First, we
would like to go beyond the calculation of purely propagative
solutions, and study the full dynamics of a given dune profile.  In
particular, as just said, an important point is the evolution of the
dome solutions when submitted to incident sand flux variations.  A
second point is to go from a 2d description -- transverse dunes -- to
real three dimensional situations.  The idea is to `cut a barchan into
longitudinal 2d slices'.  As a matter of fact, a barchan slice close
to the center of the dune looks like our dune solution, while a slice
made at the edges where the horns are present rather have a dome
shape.  Suppose these slices are completely decoupled.  Because the
small ones go faster than the large ones, an initial conical sandpile
will soon get a crescentic shape.  However, when equilibrium is
reached, all the slices should move at the same velocity.  There
should thus be a coupling between them, namely a lateral sand flux
from the centre towards the horns.  When the flux is saturated at the
crest the velocity at the crest is $c=(q_{sat}-q_0)/H$.  This suggests
than an equilibrium can indeed be achieved if $q_0$ increases in the
small slices.  Eventually the 3D dune slip face will be the sum of the
contributions of all 2D slices whose brinks depend on their heights.
Note that this scenario is consistent with the field observation that
barchan horns are more elongated at strong winds which make the
lateral sand flux less important, and consequently slices less
coupled.

Finally, quantitative comparisons between experimental dune profiles
and our theoretical predictions will be performed. This idea is to use
barchan longitudinal slices as that shown on figure \ref{ProfilTerrain},
but also sand structures under water, such as those obtained by Betat
\textit{et al}. \cite{BKFR01} or Andersen \textit{et al}. \cite{AGRL01}
for which, in principle, this model should be also valid.

\begin{acknowledgement}

\textbf{Acknowledgments}\\
We are grateful to Gerd Sauermann and Klaus Kroy for precise
explanations of all the details of their modelling.  The measurements
of the dune profiles (figure~\ref{ProfilTerrain}) were performed by B.
Andreotti, S. Douady, P. Hersen and L. Quartier.  We wish to thank
St\'ephane Douady, Christof Kr\"ulle and Brad Murray for useful discussions.
\end{acknowledgement}


\end{document}